\documentclass[prb,twocolumn,superscriptaddress, showpacs,floatfix]{revtex4-1}  
\usepackage{hyperref}
\usepackage{amssymb}
\usepackage{amsmath}
\usepackage{float}
\usepackage[usenames]{color}
\usepackage{graphicx}

\begin{document}

\title{Truncated Configuration Interaction expansions as solvers for correlated quantum impurity models and dynamical mean field theory}

\author{Dominika Zgid}
\affiliation{Department of Chemistry and Chemical Biology, Cornell University, Ithaca, New York, USA}
\author{Emanuel Gull}
\affiliation{Max Planck Institut f\"{u}r Physik komplexer Systeme, Dresden, Germany}
\author{Garnet Chan}
\affiliation{Department of Chemistry and Chemical Biology, Cornell University, Ithaca, New York, USA}

\begin{abstract}
The development of polynomial cost solvers for correlated quantum impurity models, with controllable errors, 
is a central challenge in quantum many-body physics, where these models find applications ranging from nano-science to the dynamical mean-field theory (DMFT). Here we describe how 
 configuration interaction (CI) approximations to exact diagonalization (ED)
may be used as solvers in DMFT. CI approximations retain the main advantages of ED, such as the ability to treat
general interactions and off-diagonal hybridizations and to obtain real spectral information, but are of polynomial cost. Furthermore,
their errors can be controlled by monitoring the convergence of physical quantities as a function of the CI hierarchy.
Using benchmark DMFT applications, such as single-site DMFT of the 1D Hubbard model and  $2\times 2$ cluster DMFT of the 2D Hubbard model,
we show that CI approximations allow us to obtain near-exact ED results for a tiny fraction of the cost.
This is true over the entire range of interaction strengths including ``difficult'' regimes, such as in the pseudogap phase of the 2D Hubbard model. 
We use the ability of CI approximations to treat large numbers of orbitals to 
 demonstrate convergence of the bath representation in the $2\times 2$ cluster DMFT using a 24 bath orbital representation.
CI approximations thus form a promising route to extend ED to problems
that are currently difficult to study using other solvers  such 
as continuous-time quantum Monte Carlo, including  impurity models with large numbers of orbitals and general interactions.
\end{abstract}


\maketitle

\section{Introduction}
Quantum impurity models
describe a finite set of interacting ``impurity'' orbitals coupled to a large number of non-interacting ``bath'' or ``lead'' states.
They were originally designed to describe the effect of magnetic impurities embedded in a non-magnetic host material\cite{Anderson61}, 
but have since found a wide variety of applications ranging from nanoscience,\cite{Hanson07} where they are used to describe quantum dots and molecular conductors, to surface science\cite{Brako81}, for the description of molecule adsorption on a substrate, to research in quantum field theories.\cite{Wilson75,Affleck08}
In recent years, they have gained an increasingly important role in condensed matter and materials science, where they appear as auxiliary models in the simulation of correlated lattice models within the so-called dynamical mean field theory (DMFT)\cite{Metzner89,Georges92,Georges96} approximation and its extensions.\cite{Kotliar06,Maier05}

A general quantum impurity model is described by the Hamiltonian
\begin{align}
H&=H_\text{loc}+H_\text{bath}+H_\text{hyb}, \\
\label{HQIM}
H_\text{loc}  &= \sum_{pq} t_{pq}d^\dagger_p d_q +\sum_{pqrs}I_{pqrs}d^\dagger_p d^\dagger_q d_r d_s, \\
H_\text{bath} &=\sum_{k i} \varepsilon_{k i}c^\dagger_{k i}c_{k i},\\
H_\text{hyb}  &=\sum_{k i p}{V}_{kip}c^\dagger_{ki}d_p +\text{h.c.}.
\end{align}
$H_\text{loc}$ describes the ``impurity'' itself, $H_\text{bath}$ a set of non-interacting ``bath'' or ``lead'' sites, 
and the impurity-bath coupling or ``hybridization'' is contained in $H_\text{hyb}$. The operators $d^{(\dagger)}$ and  $c^{(\dagger)}$ create and annihilate impurity and lead electrons, $t$ and $I$ describe the impurity hopping and interaction terms, $\varepsilon$ a bath dispersion and $V$ the impurity-bath hybridization strength.

The finite number of impurity interactions makes quantum impurity models numerically tractable. 
The development of accurate and reliable numerical solvers for correlated quantum impurity models
 is therefore one of the central challenges of computational many-body physics.
Many different approaches have been proposed. 
Among those that can be made exact with sufficient computational effort, at least for some classes of models, are: 
quantum Monte Carlo methods, such as the continuous-time quantum Monte Carlo (CT-QMC);\cite{GullRMP}
renormalization group (RG) methods, including numerical \cite{Bulla98,Bulla08b} and density matrix RG\cite{Garcia04,Nishimoto06}; and exact diagonalization (ED)\cite{Caffarel94}.

All these techniques have different strengths and weaknesses.
For example, CT-QMC is formulated in imaginary time and real-frequency data at high frequencies, obtained with analytic continuation, is notoriously unreliable,  while
NRG has limited resolution in spectral quantities far from the Fermi surface and cannot be reliably extended beyond two impurity orbitals. ED
does not suffer from the above two difficulties, but introduces a finite size error associated with a discrete bath representation. For some special
Hamiltonians, such as those with  density-density interactions and diagonal hybridizations, 
CT-QMC has no sign problem, and thus affords a polynomial time solution of the impurity problem. However, 
for general Hamiltonians, all the above techniques including CT-QMC exhibit an exponential scaling with the number of impurity orbitals and, in the case of  ED, with the number of bath orbitals.  Consequently, there is an urgent need to develop 
controlled approximate solvers for general impurity models, where the exponential scaling is  ameliorated or eliminated.

The dynamical mean-field theory and its cluster variants provide an ideal test bed for numerical quantum impurity solvers. DMFT is now established as a powerful theoretical framework for describing interacting quantum solids, 
both in the context of single-site multi-orbital and single-orbital cluster model Hamiltonians, as well as with realistic interactions within the DFT+DMFT\cite{Kotliar06,Held06} framework.
In DMFT the bulk quantum problem is mapped onto a self-consistent quantum impurity model. Depending on the lattice model parameters, regions of weak, intermediate, and large correlation strengths can be accessed, and the wealth of previously computed data and the well-understood physics makes reliable comparisons possible.

Our present work presents controlled polynomial cost approximations to ED, 
using the idea of configuration
interaction (CI) \cite{Helgaker00} that has long been studied in quantum chemistry. Recall that, at zero temperature,
the Green's function is
\begin{align}
i g_{ij}(\omega) &= \langle \Psi| a^\dag_i (\omega - H + E - i\eta)^{-1}  a_j|\Psi \rangle \nonumber \\ &+ 
\langle \Psi | a_j (\omega + H-E +i\eta)^{-1} a^\dag_i |\Psi\rangle \label{eq:greensfn}
\end{align} 
where $E$ and $\Psi$ are the ground-state energy and wavefunction of $H$. In ED, the true ground state wavefunction $\Psi$
is expanded in the complete space of Slater determinants, and the size of the complete space, which scales exponentially 
as a function of the number of impurity and bath orbitals, is the primary limitation of the calculation.
Even using state-of-the-art ED (Lanczos) codes, no more than 16 electrons in 16 orbitals (32 spin-orbitals) can be treated. CI
approximates ED by solving for $\Psi$ within a {\it restricted} variational space of Slater determinants. 
This variational space is constructed by including determinants based on their excitation level relative to a single,
or multiple, physically motivated reference determinants. 
The various CI methods form a convergent hierarchy of approximations, where the variational space
is systematically increased, and thus their error, relative to the theoretical ED limit,  can be controlled by monitoring  the convergence of the hierarchy. 
Furthermore, because CI methods exhibit a polynomial
scaling with respect to the number of impurity and bath orbitals, they have
the  potential to treat  much larger systems than ED.
Indeed, in quantum chemistry, CI calculations
with a thousand orbitals are routine. 

The central question to answer in the context of correlated quantum impurity models is
whether or not CI approximations form a sufficiently rapidly convergent hierarchy for the
physical quantities of primary interest. If so, the  ability to treat large numbers of orbitals and off-diagonal hybridizations, while retaining the strengths of ED,
can be expected to be of great utility in revealing the physics of complex quantum impurity models.
In Ref.~\onlinecite{Zgid11} we demonstrated that a very approximate CI solver
could reproduce exact diagonalization results in a simple quantum impurity problem arising from the DMFT approximation to the cubic hydrogen solid.
However, in that work our focus was not on the quality of the solver, but rather on  chemical aspects of DMFT, such
as the use of realistic Hamiltonians which do not suffer from double-counting. 
In the current work, we return to a systematic study of the CI approximations themselves.
Since our target is to assess the quality of our approximations, we concentrate here on well-studied
DMFT benchmark problems whose physics is understood, including
 single-site DMFT of the 1D Hubbard model and 2$\times$2 cluster DMFT
of the 2D Hubbard model. The double-counting issue does not arise in these systems.
As we will demonstrate, in these systems the  CI approximations
allow us to reproduce the ED calculations at a small fraction of the cost. Furthermore, because we can treat a larger number of orbitals than with ED, we will demonstrate that we can converge these models
with respect to their bath representation. In the cases of the 2$\times$2 cluster DMFT approximation to the Hubbard model this
has previously not been possible with ED (Lanczos).

The structure of this paper is as follows. In section \ref{sec:methsec} we first describe the theory behind CI approximations,
including a detailed description of the excitation space, single- and multi-reference CI approximations,
complete active spaces, and natural orbitals. In section \ref{sec:impsec}, we briefly describe some technical details of the implementation.
In section \ref{sec:results} we describe our application of CI solvers in model DMFT problems described above, using ED as a comparison
where possible, and we demonstrate further the ability of CI to converge systems with large numbers of bath orbitals. Finally, we describe
perspectives and conclusions in section \ref{sec:conclusions}.

\section{Configuration interaction approximations} \label{sec:methsec}

Configuration interaction (CI) wave functions $|\Psi_\text{CI}\rangle$ are a set of systematic approximations to the ED wave function $\Psi_\text{ED}$. Using CI wave functions, the ground state of the impurity model is determined in a truncated subset of the complete set of Slater determinants.
Once a ground state wave function is obtained, the impurity Green's function and self-energy, the central
quantities in DMFT, are evaluated through Eq.~(\ref{eq:greensfn}).
 CI truncations rely on an {\it a priori} ranking of the importance of the determinants, in terms of excitation character 
relative to a single starting determinant (single-reference CI), or multiple starting determinants (multi-reference CI). 
This ranking is motivated by ordinary and degenerate perturbation theory, although CI approximations are not perturbative approximations {\it per se}.
Note that the accuracy of CI truncations of the determinant space depends on the choice of orbital basis, and this is also an important consideration in a CI calculation.


\begin{figure}[htb]
\begin{center}
 \includegraphics[width=0.8\columnwidth]{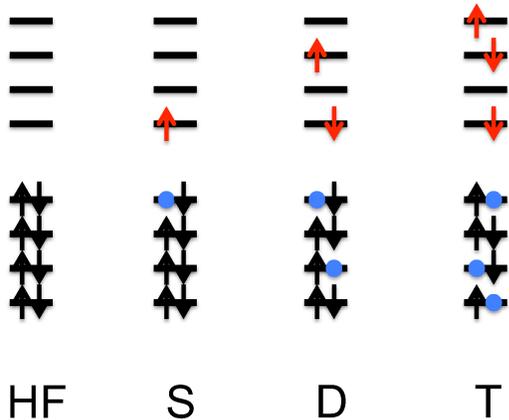} 
\caption{Schematic of determinants included in configuration interaction approximations. HF denotes
the Hartree-Fock determinant with a set of doubly occupied orbitals. S denotes a singles excitation, where
one particle (arrow, red online) is excited out of a doubly occupied orbital (leaving a hole, dot, blue online). D and T denote
doubly and triply excited configurations with respect to the Hartree Fock reference state. \label{fig:ci}}
\end{center}
\end{figure}

\begin{figure*}[htb]
\begin{center}
\begin{tabular}{cc}
(a) \includegraphics[width=0.8\columnwidth]{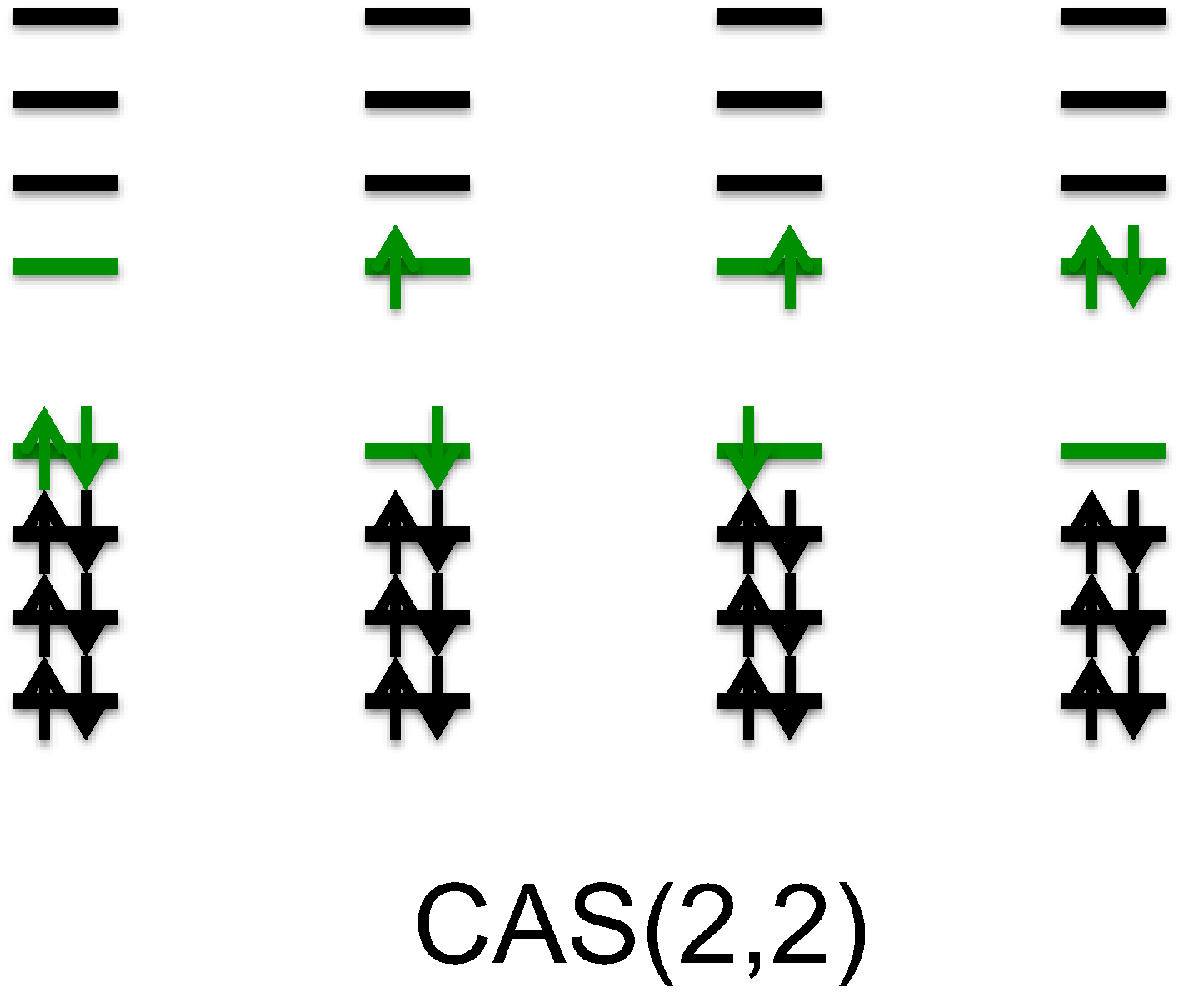}  & \hspace{2cm}
(b) \includegraphics[width=0.8\columnwidth]{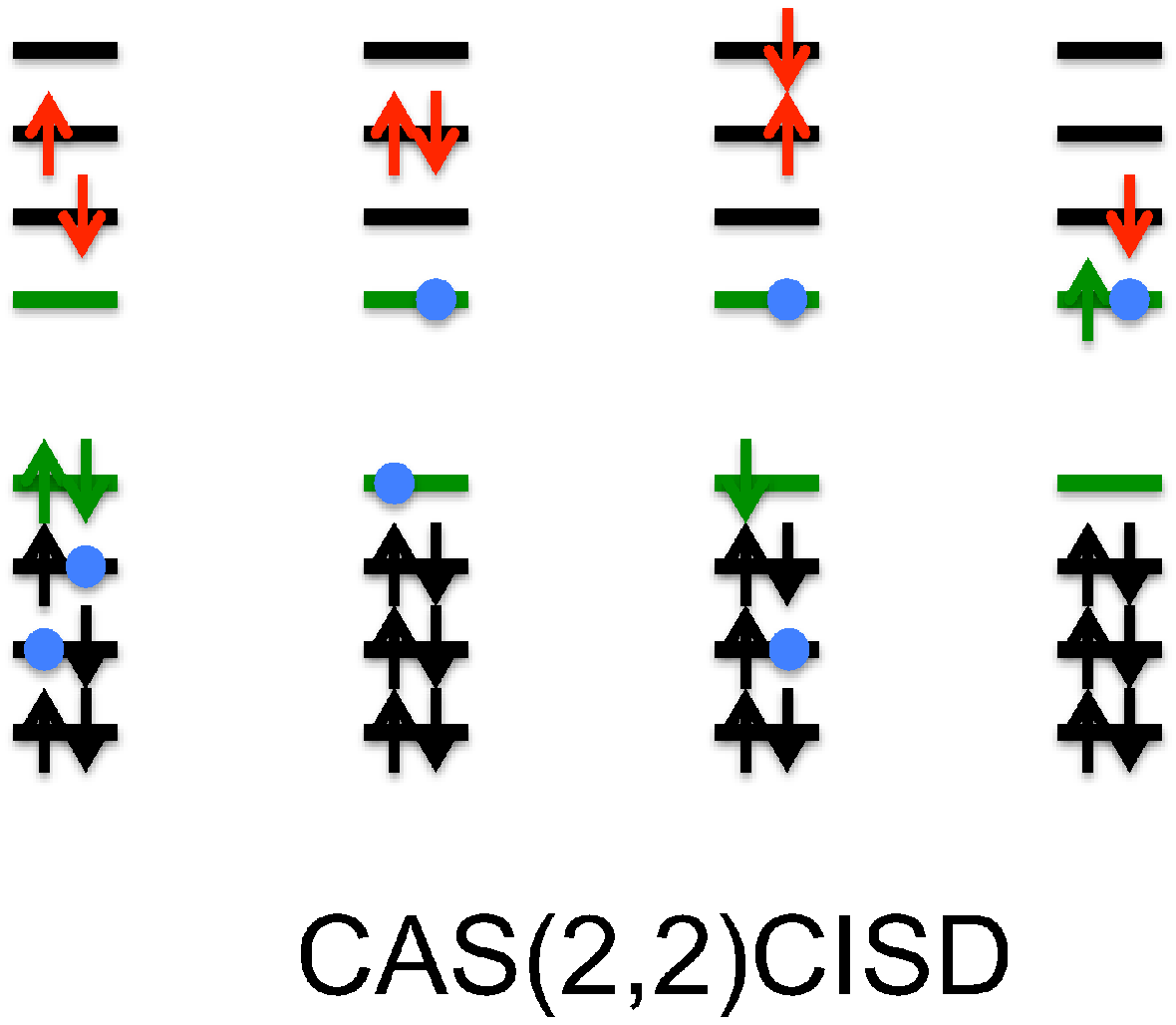}  
\end{tabular}
\caption{(a): determinants in a CAS(2,2) complete active space. The light (green online) lines are the 2 ``active'' orbitals; the rest
are denoted ``inactive''. 
The four configurations correspond to the four ways of distributing 2 electrons across the 2 active orbitals. (b): 
examples of four of the determinants contained in a CAS(2,2)CISD approximation. Excitations may be from doubly occupied, non-active orbitals (first determinant),
from active orbitals (second, fourth determinants), and from a mixture of active and inactive orbitals (third determinant). \label{fig:casci}}
\end{center}
\end{figure*}

We first motivate single-reference CI approximations using the Anderson model
in the limit of small $U$. Here, $\Psi$ is close to the 
Hartree-Fock (HF) determinant $\Phi$. Consequently, we take $\Phi$ to be the (single) reference determinant in the CI.
We next change from the site basis ($d, d^\dag, c, c^\dag$) to the basis of 
HF orbitals ($a, a^\dag$), which are a mixture of impurity and bath orbitals. 
The determinants in the HF basis can be labeled in terms of excitation or particle-hole
character relative to $\Phi$. For example, a singly-excited determinant $\Phi_i^a=a^\dag_a a_i (\Phi)$
has one particle and one hole relative to the HF determinant; the doubly-excited determinant $\Phi_{ij}^{ab}=a^\dag_a a^\dag_b a_i a_j (\Phi)$
has two particles and holes, and so forth. 
To construct a CI approximation,  we truncate the complete  determinant space
 based on the maximum excitation character
of determinants in the expansion of $\Psi$. For example, in a  CI singles and doubles (CISD) approximation,
we approximate $\Psi$ with an expansion with at most doubly excited determinants
(see Fig. \ref{fig:ci}), 
\begin{align}
|\Psi\rangle \approx c_0 |\Phi\rangle + \sum_{i,a} c_i^a |\Phi_i^a\rangle + \sum_{ij,ab} c_{ij}^{ab} |\Phi_{ij}^{ab}\rangle.
\end{align}
More accurate CI approximations, with up to triple (CISDT), quadruple (CISDTQ) and higher excitations
can be formulated in a similar way, and these
systematically approach the full ED solution.
However, for small $U$, we can expect CISD to already be a good approximation, because it contains all classes of determinants
that couple with  $\Phi$ through first-order perturbation theory in $U$. Similarly, CISDTQ contains
all classes of determinants that couple through second-order, and so on.

The above CI approximations (CISD, CISDT, etc.) are termed single-reference because
the truncation of the  determinant space is based on excitations relative to a single reference determinant $\Phi$.
We expect this hierarchy of truncations to be rapidly convergent for small $U$, but for large $U$ 
multiple determinants can become degenerate with $\Phi$ on the scale of $U$ and contribute
with similar weights to $\Psi$. For example, in the single site Anderson model  at large $U$, $\Psi$
is qualitatively given by a superposition of two determinants describing a ``Kondo'' singlet
coupling of electrons of opposite spin on the impurity orbital and in the bath.
In such cases, it is more reasonable to rank determinants with respect
to a {\it set} of near-degenerate determinants. This is the basis of
 multi-reference CI approximations. 
Denoting  the near-degenerate determinants as $ \Phi(I)$ (where $I$ ranges
over the degenerate set), then for each $\Phi(I)$, we can
define singly, doubly, and higher excited determinants, $\Phi_i^a(I)=a^\dag_a a_i (\Phi(I))$, $\Phi_{ij}^{ab}(I)=a^\dag_a a^\dag_b a_i a_j (\Phi(I))$, and so on.
In the multi-reference CI singles doubles approximation,  $\Psi$ is expanded in
\begin{align}
|\Psi\rangle &\approx \sum_I c_0(I) |\Phi(I)\rangle + \sum_{I} \sum_{i,a} c_i^a(I) |\Phi_i^a(I)\rangle \\ \nonumber &+ \sum_{I}\sum_{ij,ab} c_{ij}^{ab}(I)|\Phi_{ij}^{ab}(I)\rangle.  \label{eq:casci}
\end{align}
Multi-reference approximations including triples and higher excitations can be defined analogously.

One drawback of multi-reference CI calculations
is that they are more difficult to set up and  describe compactly, because of the need to 
 identify
the near-degenerate set of determinants $ \Phi(I)$. 
A much simpler task is to specify only a set of near-degenerate {\it orbitals}, 
and  to assume that all determinants  obtained by considering  
 different occupancies of the near-degenerate orbitals (with
the remaining orbitals held empty or doubly occupied)
constitute a near-degenerate set of references.
 This
is the basis of the complete active space (CAS) specification of the references $\Phi(I)$. A CAS$(n,m)$ set of references
is obtained by identifying $m$ near-degenerate orbitals, and constructing
the determinants consisting of all distributions of $n$ particles across the $m$ orbitals.
 Once the CAS space  is constructed, we can define a CASCI as above.
For example, the CASCISD approximation consists of expanding $\Psi$ using a space of singles and doubles excitations out of the CAS$(n,m)$ space as in Eq.~(\ref{eq:casci})
(see Fig.~\ref{fig:casci}), and higher  analogues are similarly defined. In this work, we will exclusively use CAS
spaces when defining our multi-reference CI calculations.

\subsection{Improved orbitals}\label{subsec:orbital}

It is clear that the accuracy of a CI approximation
is dependent on the orbitals used to specify the determinants. For small $U$, the HF orbital basis
performs well. However, this is not always the best
 choice when $U$ is large. The orbital basis which leads
to the most rapid convergence of a complete CI expansion, as measured by one-particle
density matrix norm,  is called the {\it natural orbital} basis.
Natural orbitals are eigenfunctions of the one-particle density matrix
 $D_{ij} = \langle \Psi | a^\dag_i a_j |\Psi\rangle$, where $\Psi$ is the exact or CI wavefunction; the corresponding eigenvalues are the natural occupancies.
The natural orbital basis is a commonly used basis for CI calculations.
Of course, the natural orbitals themselves are defined using $\Psi$ which
is not known until the CI calculation is performed. Consequently, a natural orbital based CI calculation
is usually carried out in two steps.\cite{ino} First, a CI calculation in the Hartree-Fock basis
is performed to obtain the density matrix. This is then diagonalized to obtain the natural orbitals, and the CI calculation 
is repeated in this basis. In principle, this procedure can be iterated, although we have not done so in the calculations in this work.

These procedures are standard in modern quantum chemistry and are commonly used to treat finite molecular systems. Detailed descriptions can be found in Refs.~\onlinecite{Helgaker00,Sherrill99}.

\section{Implementation}\label{sec:impsec}

We have implemented the CI based approximations
described above, as well as ED, within the context of single-site
and cluster-based DMFT solvers.
Our code uses  a modified version of the efficient
string-based CI algorithm in the \textsc{Dalton} 
quantum chemistry package.\cite{Dalton05} 
The solution of the CI and ED eigenvalue problems is carried out using iterative
Davidson diagonalization,\cite{Davidson75} while the determination of the Green's function in Eq.~(\ref{eq:greensfn})
is carried out using the Lanczos algorithm.\cite{Lanczos50} For the hybridization and bath fitting necessary in the DMFT context, 
we have employed the procedures described in Ref.~\onlinecite{Zgid11}. The DMFT self-consistency was carried out until convergence in the self-energy was reached with a tolerance of less than $0.5\%$.
All calculations were performed at $T=0$, and all energies are in units of $t=1$. The $\beta$ used for fitting the dynamical mean field parameters were $20/t$ (single site DMFT) and $12.5/t$ (plaquette).

\section{Results}\label{sec:results}

We now assess the performance of CI approximations as quantum impurity solvers using  established
benchmark problems. To recapitulate, the two central questions are:
 how rapidly do the CI approximations converge to  ED, for example, as a function of excitation level or orbital basis, and, do
CI approximations allow us to accurately treat a larger number of orbitals than ED? 
We found the impurity models occurring in the dynamical mean field context to be more difficult to solve than simple impurity sites coupled to an analytically constructed density of states, and we therefore focus our presentation on impurity models obtained within this context. We first study an impurity model without self-consistency imposed. We then examine
two  DMFT models. The first is the single site DMFT approximation to the 1D Hubbard model. Here, ED
 calculations can be  converged with respect to the number of bath orbitals, which allows us to
  compare ED and CI approximations in the limit of a converged bath representation. Our second model is a 2$\times$2 plaquette (4-site) cellular dynamical mean field\cite{Kotliar01,Maier05} calculation
for the 2D Hubbard model. Such 4-site cluster models have been extensively studied  with 
ED~\cite{Lichtenstein00,Kotliar01,Civelli05,Kancharla08,Liebsch08,Liebsch09} 
as well as with CT-QMC,\cite{Haule07,Gull08_plaquette,Park08plaquette} 
and  provide a standard calibration point.  We begin by comparing ED and CI approximations using an
ED parametrization with 8 bath orbitals. Next, we demonstrate the ability of CI methods to treat large numbers of orbitals
by converging the 4-site cluster model with respect to the number of bath orbitals in the parametrization. Our largest
calculation involves 28 orbitals, significantly larger than can be treated with ED. 
In the appendix we present 
a three-orbital single site DMFT calculation of a model relevant for the physics of the $t_{2g}$ bands in transition metal oxides. This model uses
 a Slater-Kanamori form of the impurity interaction.\cite{Mizokawa95,Imada98} We show that CI approximations can be used
with a general impurity Hamiltonian with non-density-density interactions and demonstrate convergence of the bath representation
with up to 24 orbitals.

\subsection{Anderson impurity model}
As a test case for a quantum impurity model we present in Tab.~\ref{tab:bathparm} the parametrization for typical hybridization strengths and energy level parameters as they arise in the DMFT context, for $U/t=4$ at half filling.
The Hamiltonian of this impurity model is
\begin{align}
H&=U\left(n_\uparrow n_\downarrow-\frac{n_\uparrow+n_\downarrow}{2}\right) +\sum_{i\sigma} \varepsilon_{i}c^\dagger_{i\sigma}c_{i\sigma}\\ \nonumber
&+\sum_{i\sigma}{V}_{i}c^\dagger_{i\sigma}d_{\sigma} +\text{h.c.}.
\end{align}

The impurity model has one impurity site and eleven bath sites.
In the particle-hole symmetric case, the choice of the active orbitals is motivated by the energetic degeneracy of the eigenvalues present already in the non-interacting Hamiltonian.
The active orbitals for the CAS calculation are the orbitals $6$ and $7$ of the natural orbitals displayed (as obtained in ED) in Tab.~\ref{tab:natorbocc}, which are singly occupied.
We first remark on the sizes of the CI determinant spaces and the corresponding run-times which are given in Table \ref{tab:spacetime}.
We see that all the CI approximations involve only a small fraction of the full ED determinantal space and take a much shorter
amount of time to run. All of these calculations are doable within minutes on a desktop PC.

\begin{table}
\begin{tabular}{|c|c|c|}
\hline
i&$\epsilon_i$&$V_i$\\\hline\hline
1 &0.558819356316&0.553263286885 \\
2 &-0.558819356316&0.553263286885 \\
3 &4.45759206721&0.541358378777\\
4 &-4.45759206721&0.541358378777\\
5 &-1.47891491526&0.488524003875\\
6 &1.47891491526&0.488524003875\\
7 &-0.185401954358&0.383193040171\\
8 &0.185401954358&0.383193040171\\
9 &0.0317683411165&0.23348635632\\
10 &-0.0317683411165&0.23348635632\\
11 &0.0&1.e-5\\\hline\hline
\end{tabular}
\caption{Bath parametrization for a typical impurity problem with $12$ sites ($1$ impurity site and $11$ bath sites) obtained from converging ED, for which the spectral function and impurity self energy are reproduced in Fig.~\ref{fig:QIMFig}.}
\label{tab:bathparm}
\end{table}

\begin{table}
\begin{tabular}{|l||c|c|c|c|c|c|}
\hline
orbital number&1-4&5&6&7&8&9-12\\\hline\hline
$U/t=4$&2.00&1.895 &1.020  &0.980  &0.105  &0.000\\
$U/t=6$&2.00 &1.738 &1.009 &0.991  &0.262  &0.000\\
$U/t=8$&2.00 &1.502 &1.001 &0.998 &0.498 &0.000\\
$U/t=20$&2.00&2.000 &1.000 &1.000 &0.000 &0.000\\
\hline
\end{tabular}
\caption{Orbital occupancies in the natural orbital basis, for the impurity model of Tab.~\ref{tab:bathparm}. 
The choice of the active space is motivated by the 
partially occupied natural orbitals.\label{tab:natorbocc}}
\end{table}
\begin{table}
\begin{tabular}{|c|c|c|c|}
\hline
\hline
method & space size & $t_{GS}/t_{ED_{GS}}$ & $t_{GF}/t_{ED_{GF}}$ \\
\hline
CISD & 1819 & 0.0026263 & 0.019265\\
CISDT & 18819 & 0.012848 & 0.057334\\
CAS(2,2)CISD & 6044 & 0.012242 & 0.035215\\
CAS(2,2)CISDT & 49644 & 0.12739 & 0.13240\\
\hline
ED & 853776 & 1&1\\
\hline
\hline
\end{tabular}
\caption{Size of determinant space for $12$ electrons and $12$ orbitals, and run-times
in the solution of $\Psi$, using various CI approximations and ED. $t_{GS}/t_{ED_{GS}}$ ($t_{GF}/t_{ED_{GF}}$): runtime of ground state (Green's function) calculation with respect to ED ground state (Green's function) calculation.}
\label{tab:spacetime}
\end{table}

\begin{figure}[htb]
\begin{tabular}{c}
\includegraphics[width=0.9\columnwidth]{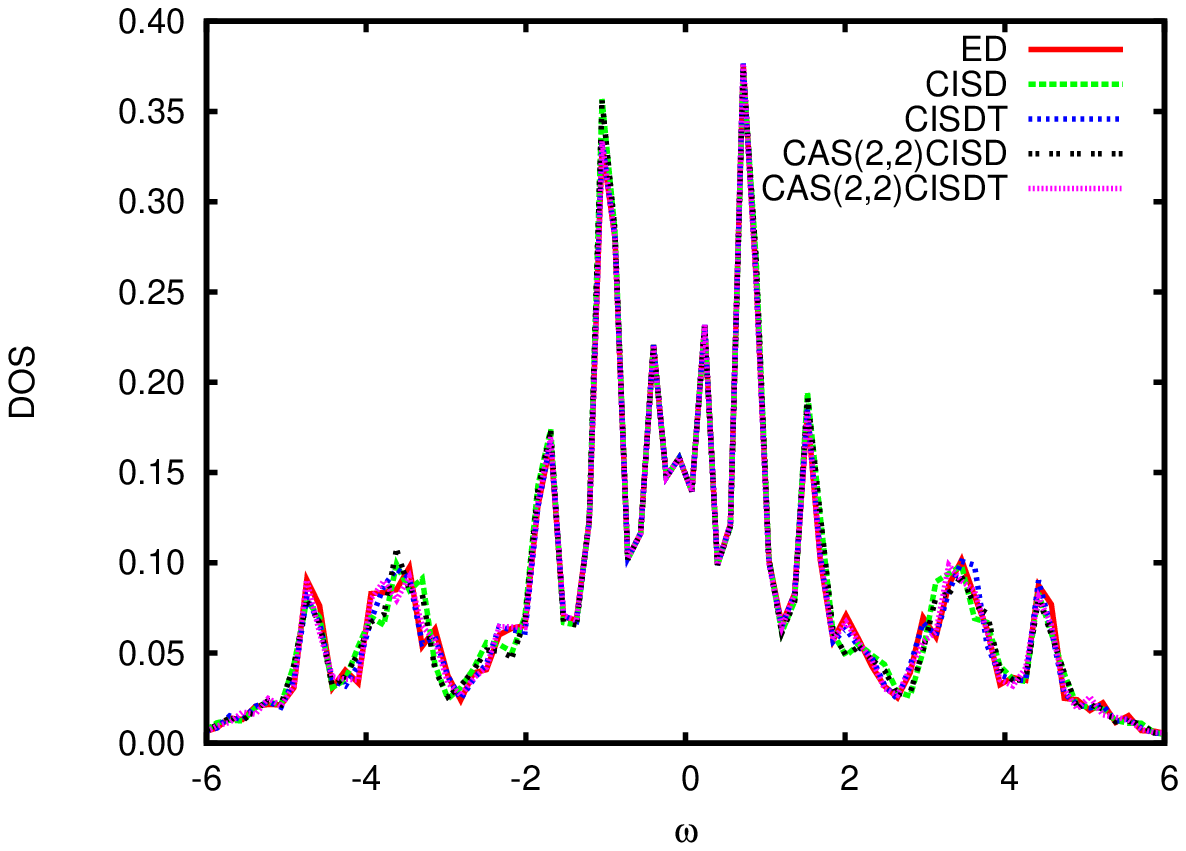}\\
\includegraphics[width=0.9\columnwidth]{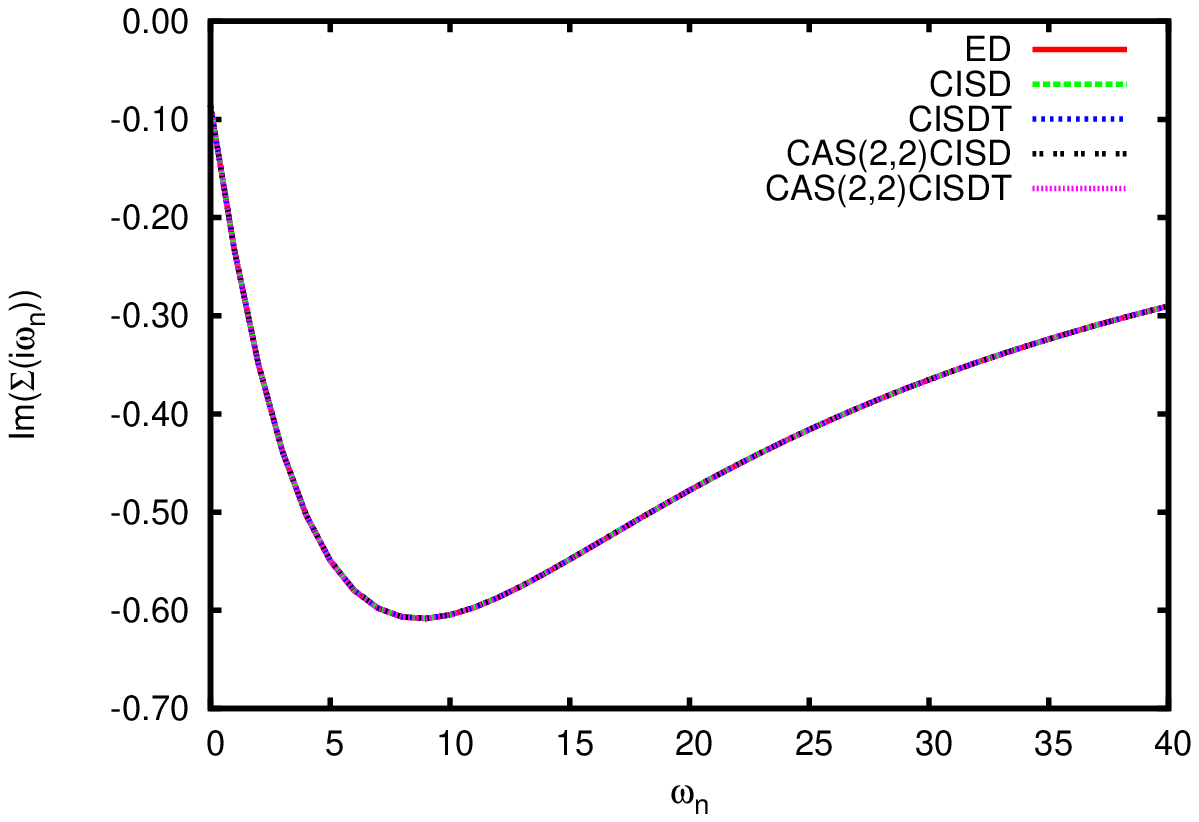}
\end{tabular}
\caption{Spectral function $-\frac{1}{\pi}\text{Im} G(\omega)$ (upper panel) and the imaginary part of the self-energy $\text{Im}\Sigma(i\omega_n)$ (lower panel)
for an impurity model using the bath parametrization in Tab.~\ref{tab:bathparm}.
Solid lines (red online): ED. Light dashed line (green online): CISD. Dark dashed line (blue online): CISDT. Double dotted line (black online): CAS(2,2)CISD. Dotted line (magenta online): CAS(2,2)CISDT.
}
\label{fig:QIMFig}
\end{figure}

Fig.~\ref{fig:QIMFig} shows results for the spectral function (upper panel) and the imaginary part of the self energy (lower panel) for the methods of Tab.~\ref{tab:spacetime}. All methods recover both the high- and the low-energy part of the self-energy to high accuracy. Differences in the spectral function are visible for $\omega > 2$, where higher excitations that are not contained within the approximations become important. Note that 
we intentionally use only a small imaginary broadening so as to preserve as much structure as possible and emphasize the difference between different approximations; This is why
 the spectral functions do not appear smooth.

Fig.~\ref{fig:QIMFig} shows that for simple Anderson impurity models the truncated CI expansions are extremely robust, and even low excitation levels can recover the proper self-energy. In the dynamical mean field context, an additional complication arises: the self-consistency condition and the bath fitting procedure lead to an amplification of differences that make the final result more sensitive to differences in the impurity self energy.

\begin{figure*}[bth]
\begin{center}
\begin{tabular}{cc}
 { (a)} \includegraphics[angle=270,width=0.95\columnwidth]{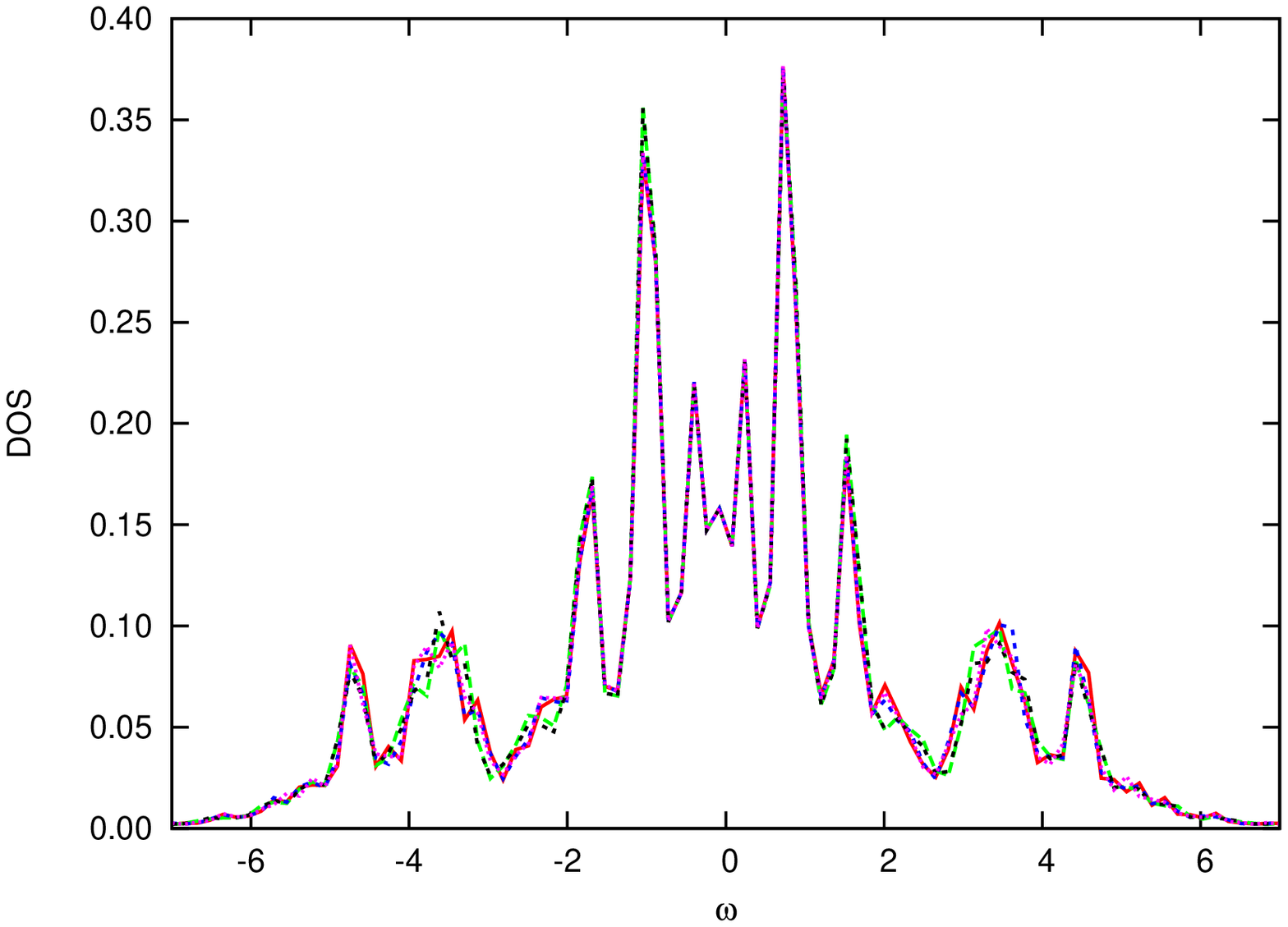} &
 { (b)} \includegraphics[angle=270,width=0.95\columnwidth]{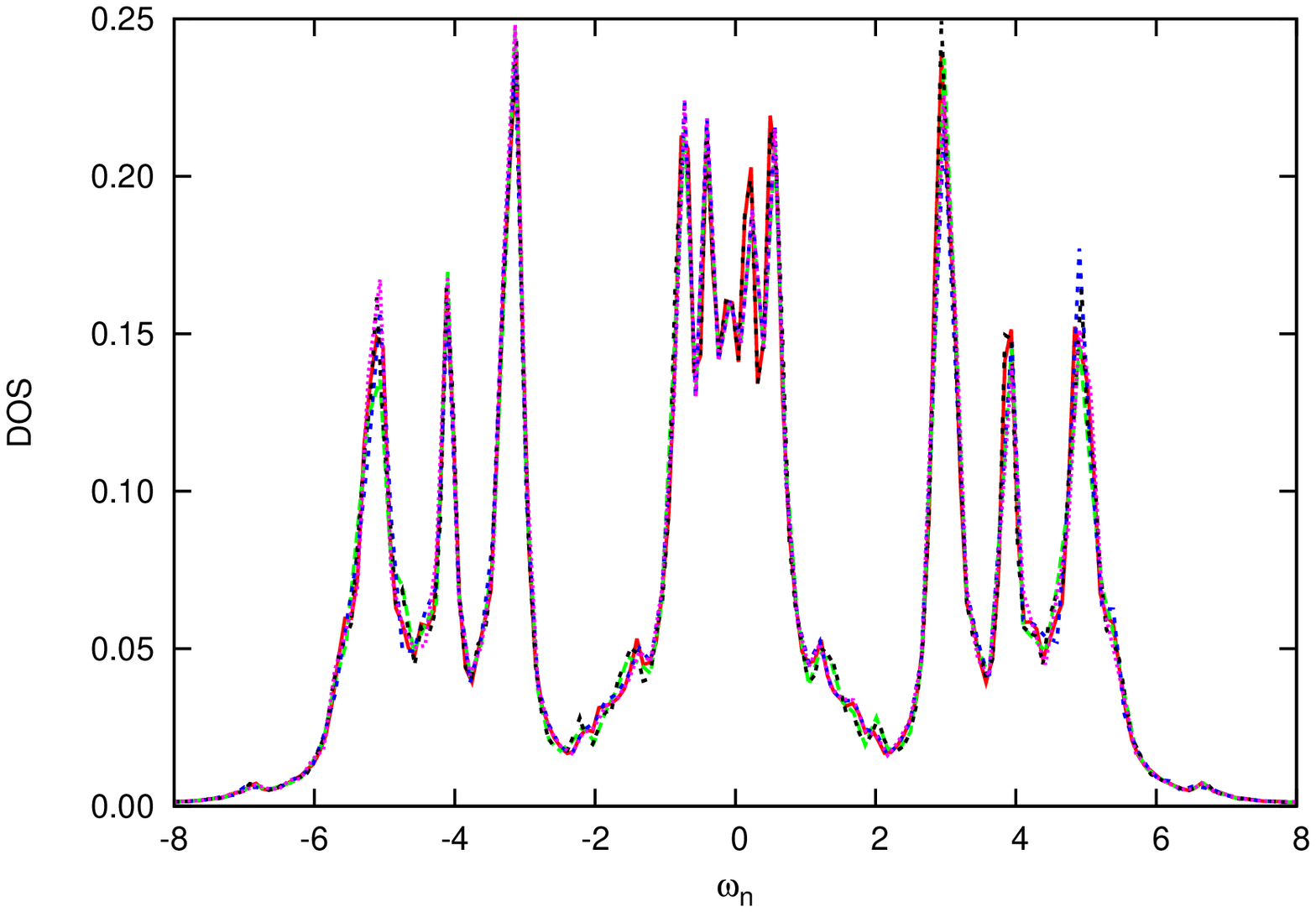} \\
 { (c)} \includegraphics[angle=270,width=0.95\columnwidth]{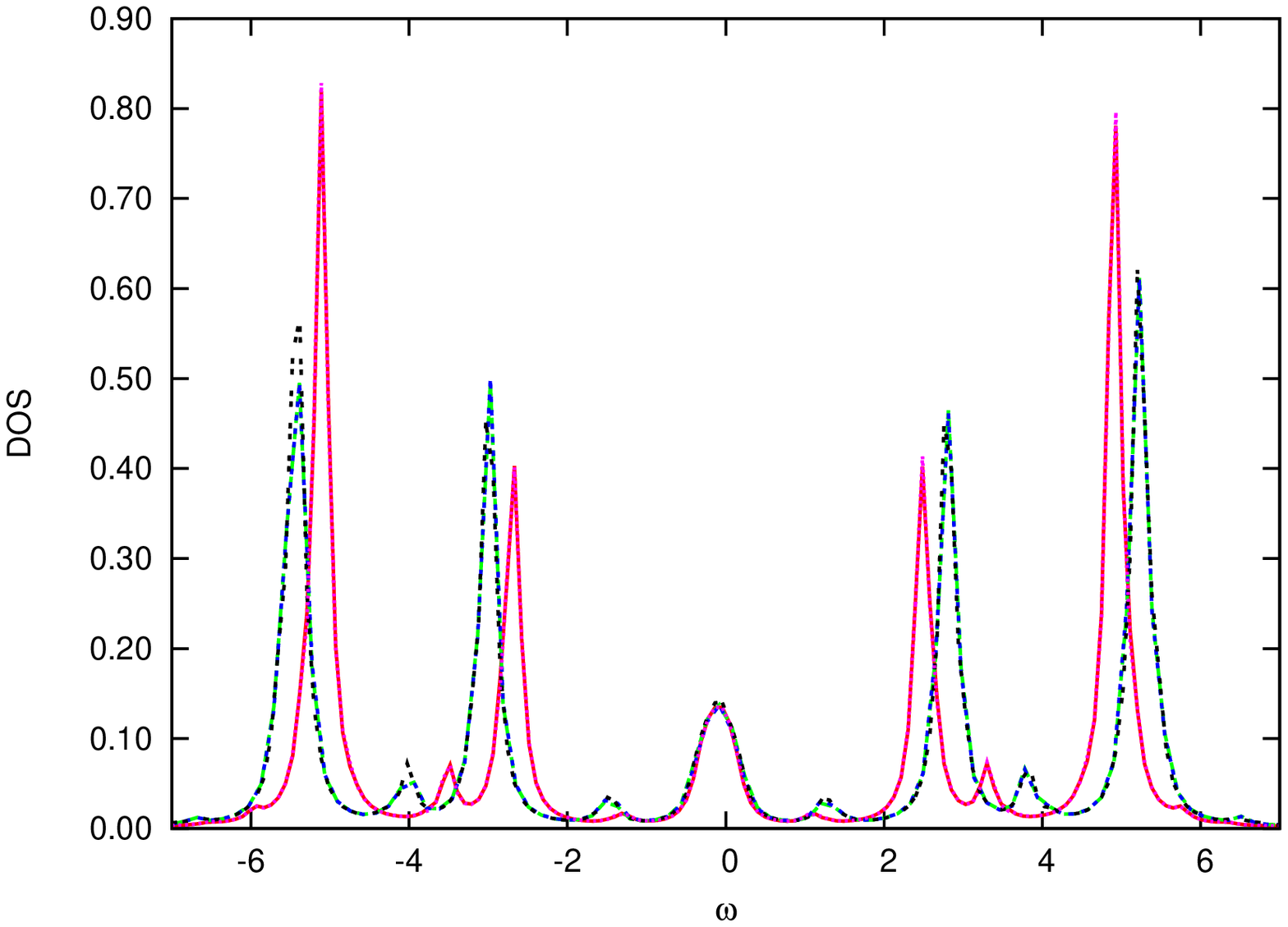} &
 { (d)} \includegraphics[angle=270,width=0.95\columnwidth]{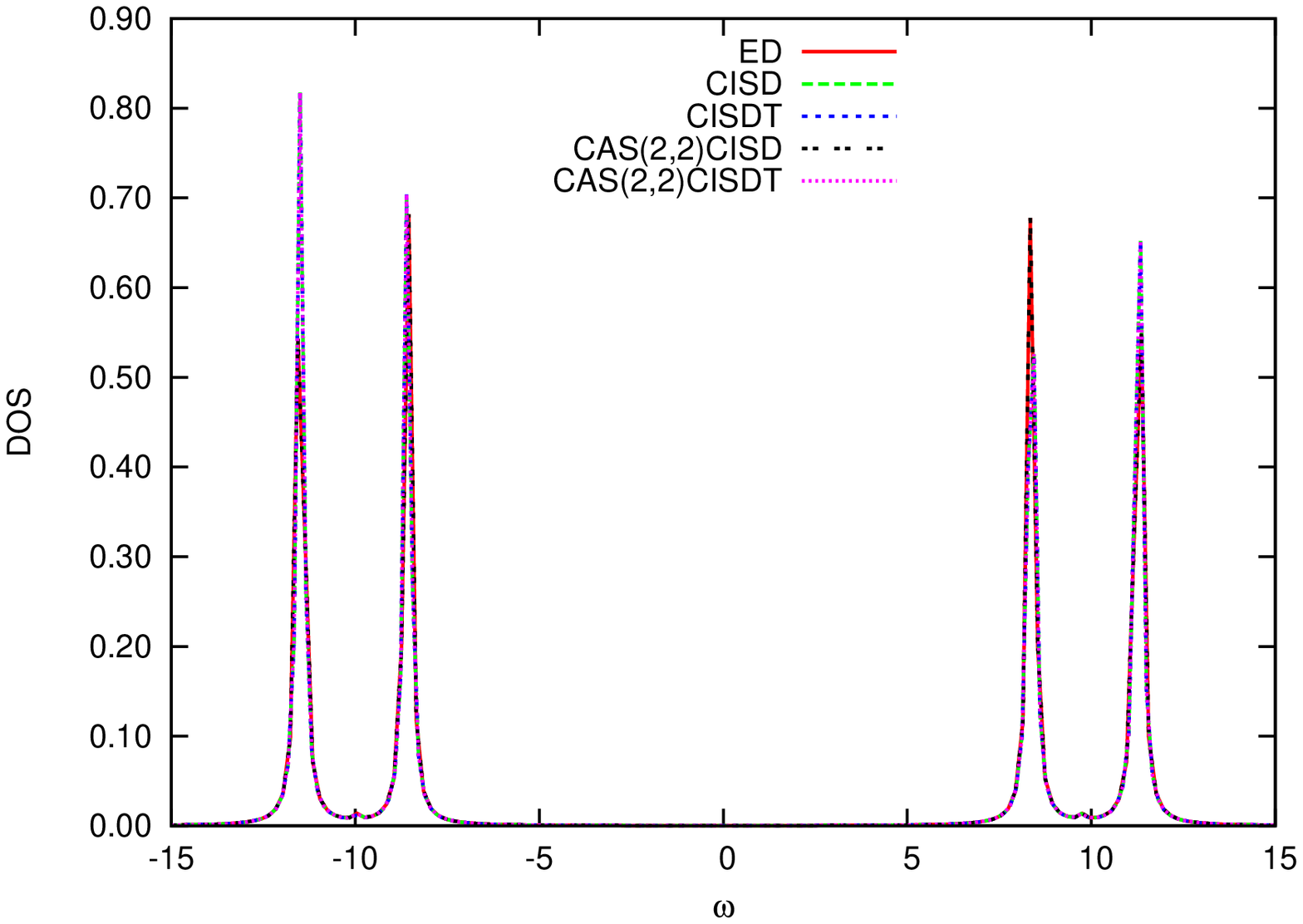} \\
\end{tabular}
\caption{Single site DMFT approximation to the 1D Hubbard model at half-filling, using $11$ bath orbitals:
Spectral function (DOS)  $A(\omega) = -\frac{1}{\pi}\text{Im} G(\omega)$. 
Solid lines (red online): ED. Light dashed line (green online): CISD. Dark dashed line (blue online): CISDT. Double dotted line (black online): CAS(2,2)CISD. Dotted line (magenta online): CAS(2,2)CISDT.
(a) $U/t=4$, (b) $U/t=6$, (c) $U/t=8$, (d) $U/t=20$. Note that panel (a) is similar (but not completely identical) to Fig.~\ref{fig:QIMFig}, where all methods used the converged ED parameters of Tab.~\ref{tab:bathparm} for solving the impurity Hamiltonian.}
\label{fig:1D_dos_hf}
\end{center}
\end{figure*}

\subsection{Single site DMFT for the 1D Hubbard model}
\subsubsection{particle-hole symmetric case}
We carried out single site DMFT calculations for the 1D Hubbard model using 
 an 11 orbital bath parametrization (12 orbitals in total).
We used CISD, CISDT, CAS$(2,2)$CISD, CAS$(2,2)$CISDT approximations as well as 
ED to obtain the spectral functions and impurity self-energies for 
 $U/t=4, 6, 8, 20$. All calculations are performed in the natural orbital basis, as described in Sec.~\ref{subsec:orbital}.

\begin{figure*}[htb]
\begin{center}
\begin{tabular}{cc}
 { (a)} \includegraphics[angle=270,width=0.95\columnwidth]{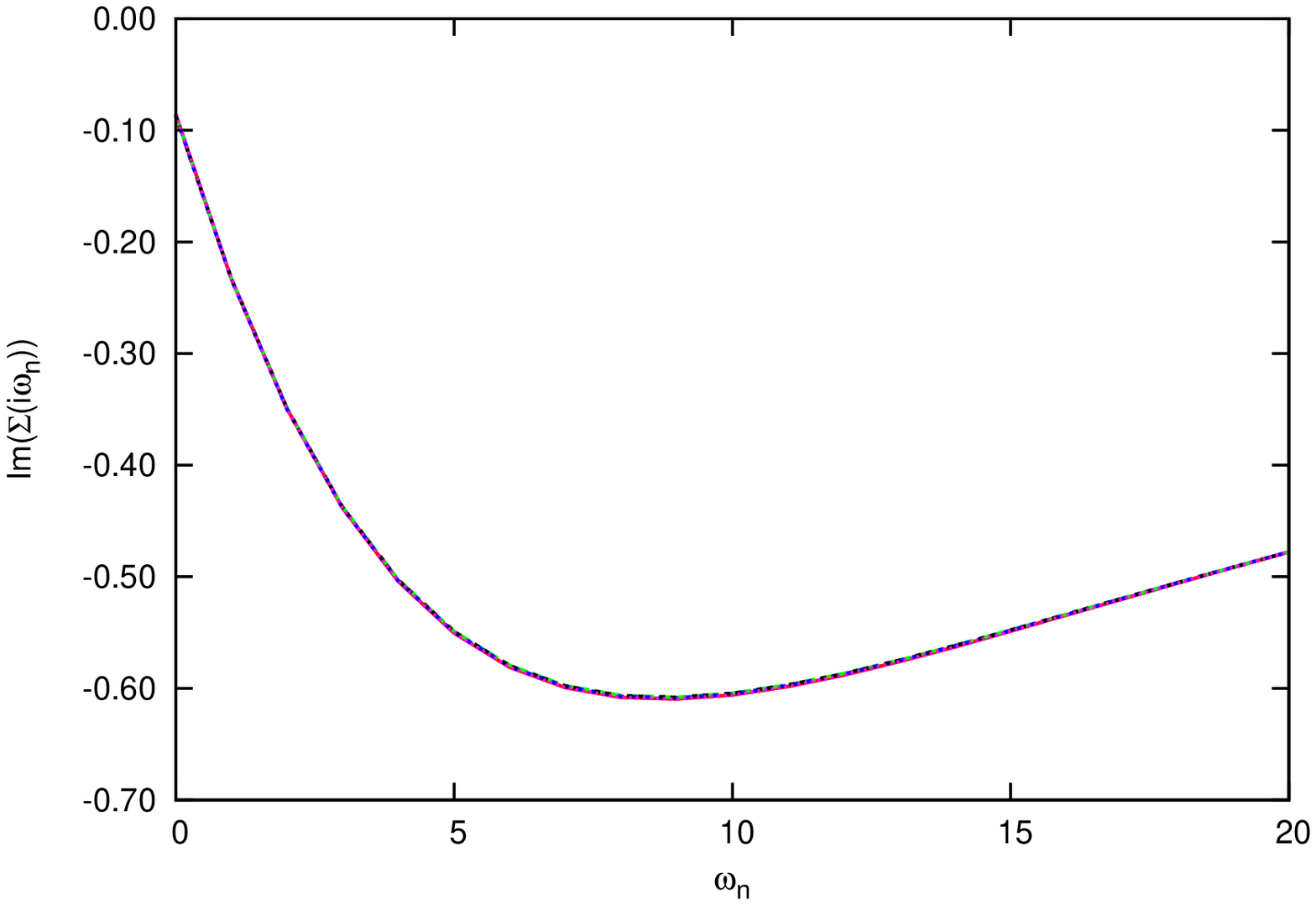} &
 { (b)} \includegraphics[angle=270,width=0.95\columnwidth]{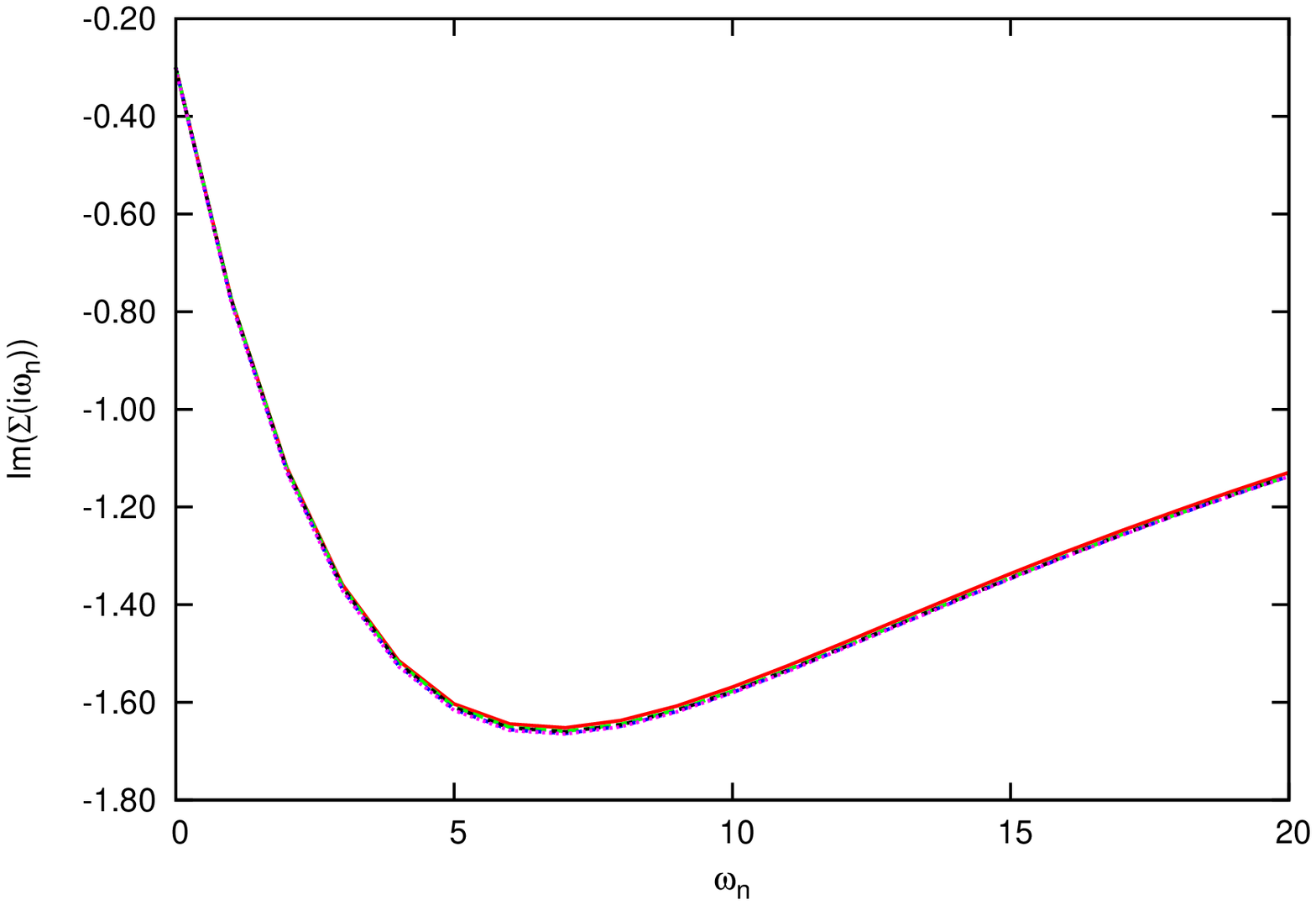} \\
 { (c)} \includegraphics[angle=270,width=0.95\columnwidth]{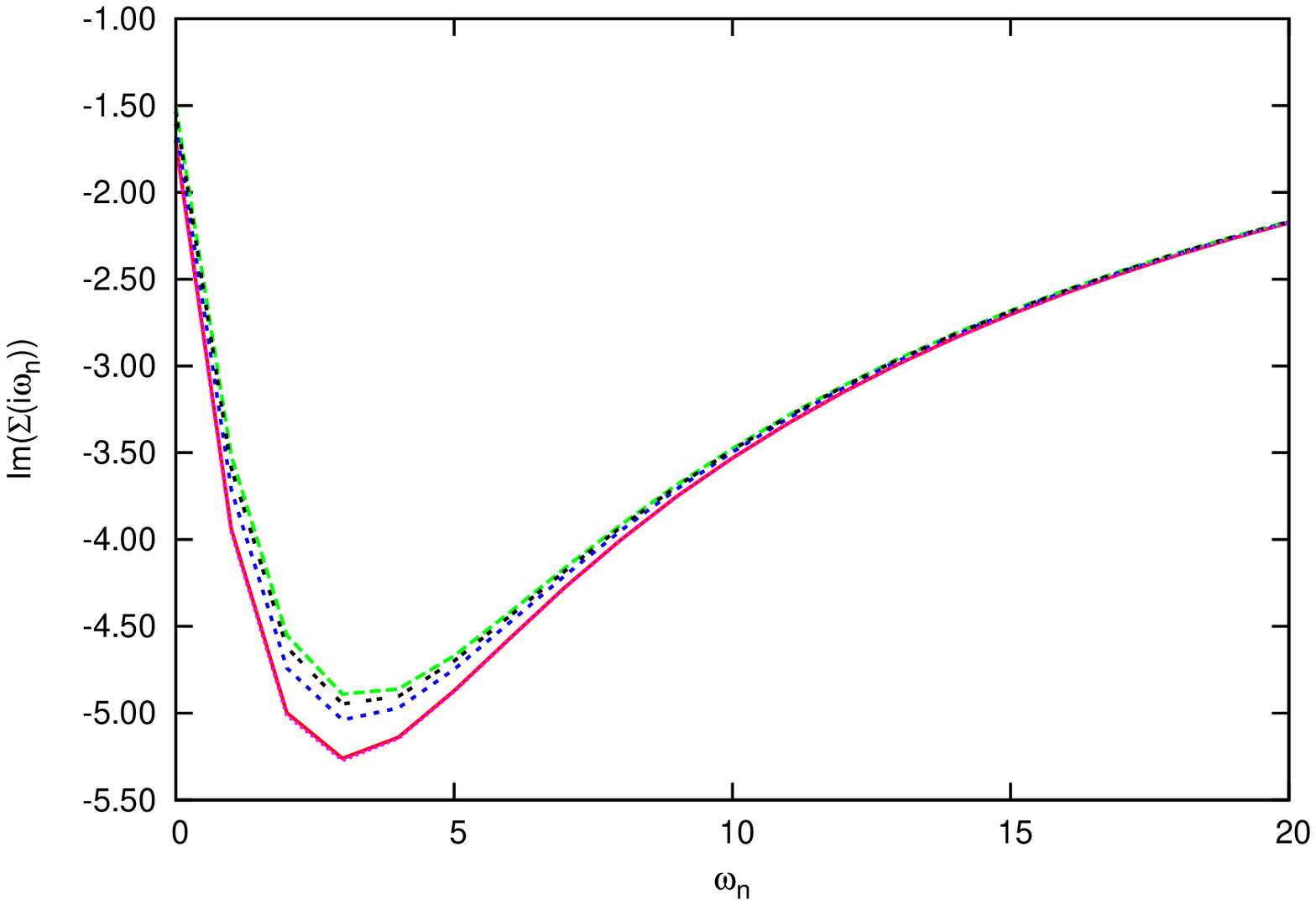} &
 { (d)} \includegraphics[angle=270,width=0.95\columnwidth]{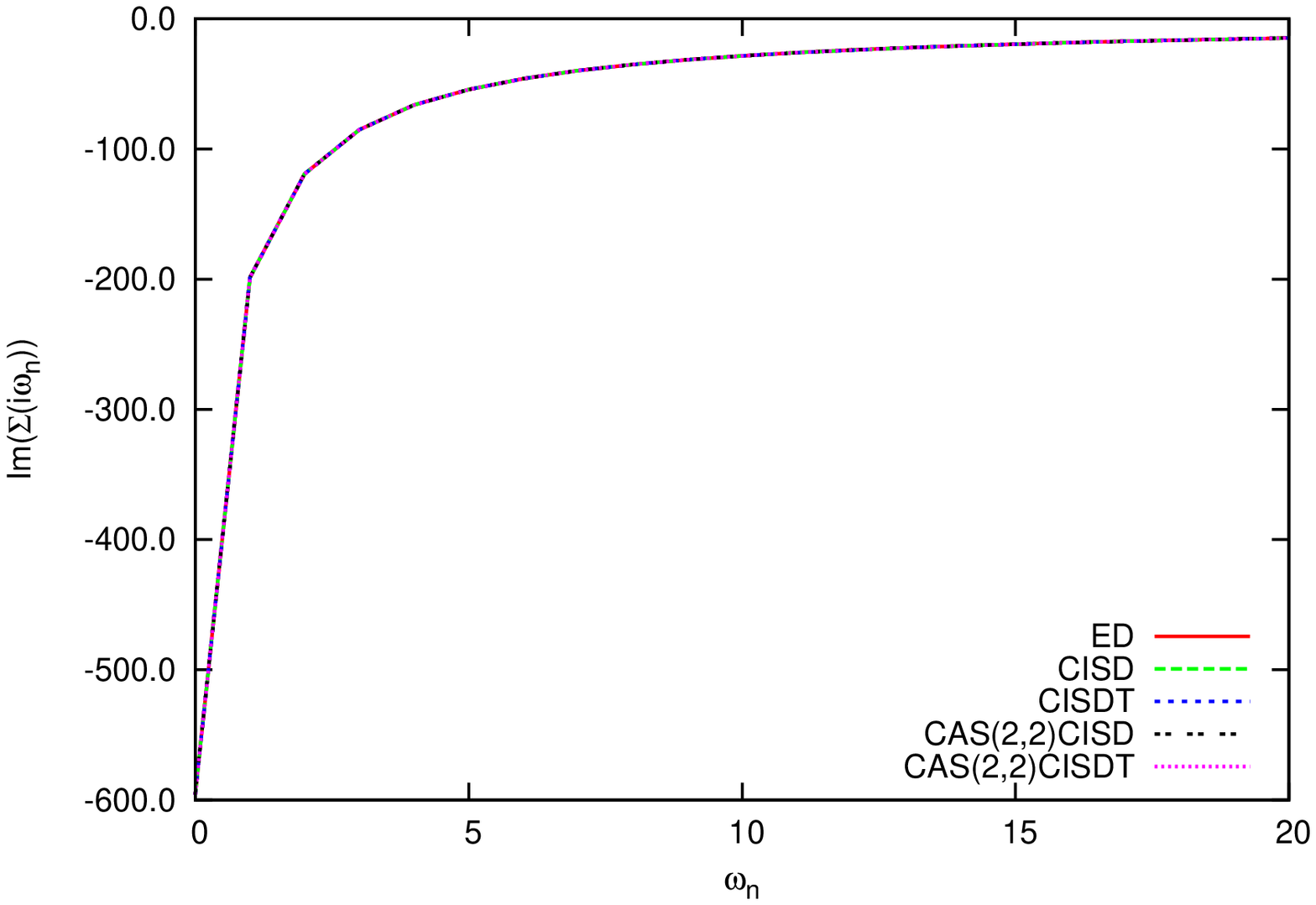} \\
\end{tabular}
\caption{single site DMFT approximation to the 1D Hubbard model at half-filling, using $11$ bath orbitals:
Imaginary part of self-energy $\text{Im} \Sigma(i\omega_n)$, with Matsubara frequencies $\omega_n = (2n+1)\pi/\beta$ for $\beta t=20$. Methods as in Fig.~\ref{fig:1D_dos_hf}.
(a) $U=4$, (b) $U=6$, (c) $U=8$, (d) $U=20$.
}
\label{fig:1s_se_error_hf}
\end{center}
\end{figure*}

The spectral functions at half-filling are shown in Fig.~\ref{fig:1D_dos_hf}.   We observe  good qualitative agreement of all CI methods with ED
 for all values of $U/t$. If we include triple excitations (CISDT or 
CAS(2,2)CISDT)  the CI spectral
functions become indistinguishable to the eye from ED. In the case of CISDT, this is achieved using
only about $2\%$ of the complete determinant space of ED. Perhaps surprisingly, multi-reference CI approximations
are not necessary  to obtain good agreement even for large $U/t$ where $\Psi$ contains large weights from
determinants other than the Hartree-Fock determinant.
 This reflects  the simplicity
of the 1D Hubbard model: the two main determinantal contributions to $\Psi$ at large $U$ differ
in the occupancies of only two electrons, which can be adequately described using doubles excitations. 
It also reflects the non-perturbative nature of CI: so long as the determinants of interest are within the CI space, 
 they can assume arbitrarily large weights in $\Psi$, and
strongly interacting (large $U$) systems can be treated.

The corresponding self-energies of the various approximations are shown in Fig.~\ref{fig:1s_se_error_hf} (we plot only the imaginary part, $\text{Im}\Sigma(i \omega_n)$).
Again, good qualitative agreement between all the CI methods and ED is observed for all values of $U/t$. Indeed, for
$U/t=4,6,20$, even the simplest CI approximation (CISD) yields an essentially indistinguishable self-energy from ED. Only at $U/t=8$ (Fig.~\ref{fig:1s_se_error_hf}c)
do we see appreciable differences. Here we need to use CAS(2,2)CISDT to achieve less than 1\% error in the self-energy. Of course,
CAS(2,2)CISDT is also the most accurate approximation to ED as measured by the size of the excitation space.

\begin{figure*}[htb]
\begin{center}
\begin{tabular}{cc}
 { (a)} \includegraphics[angle=270,width=0.95\columnwidth]{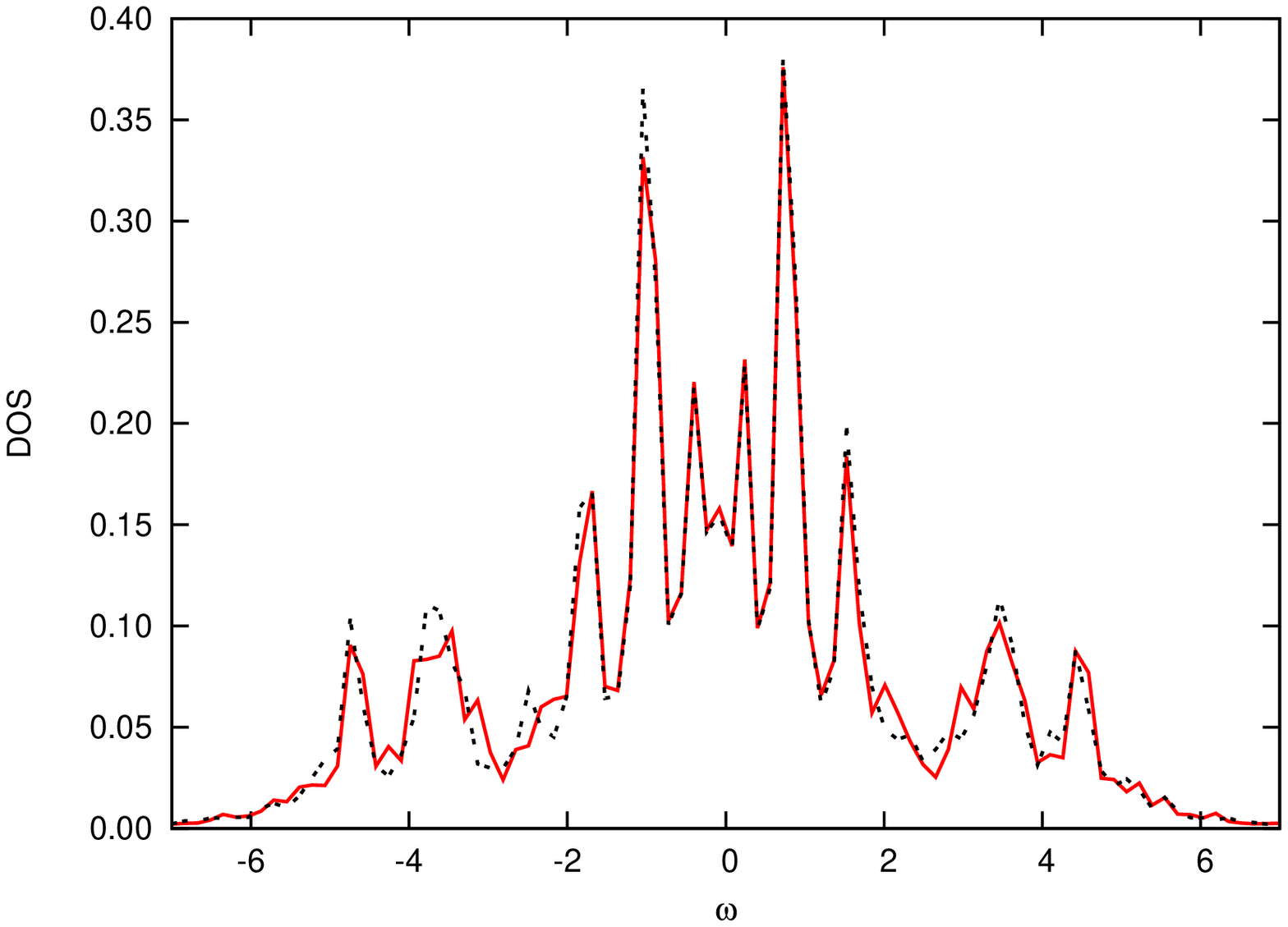} &
 { (b)} \includegraphics[angle=270,width=0.95\columnwidth]{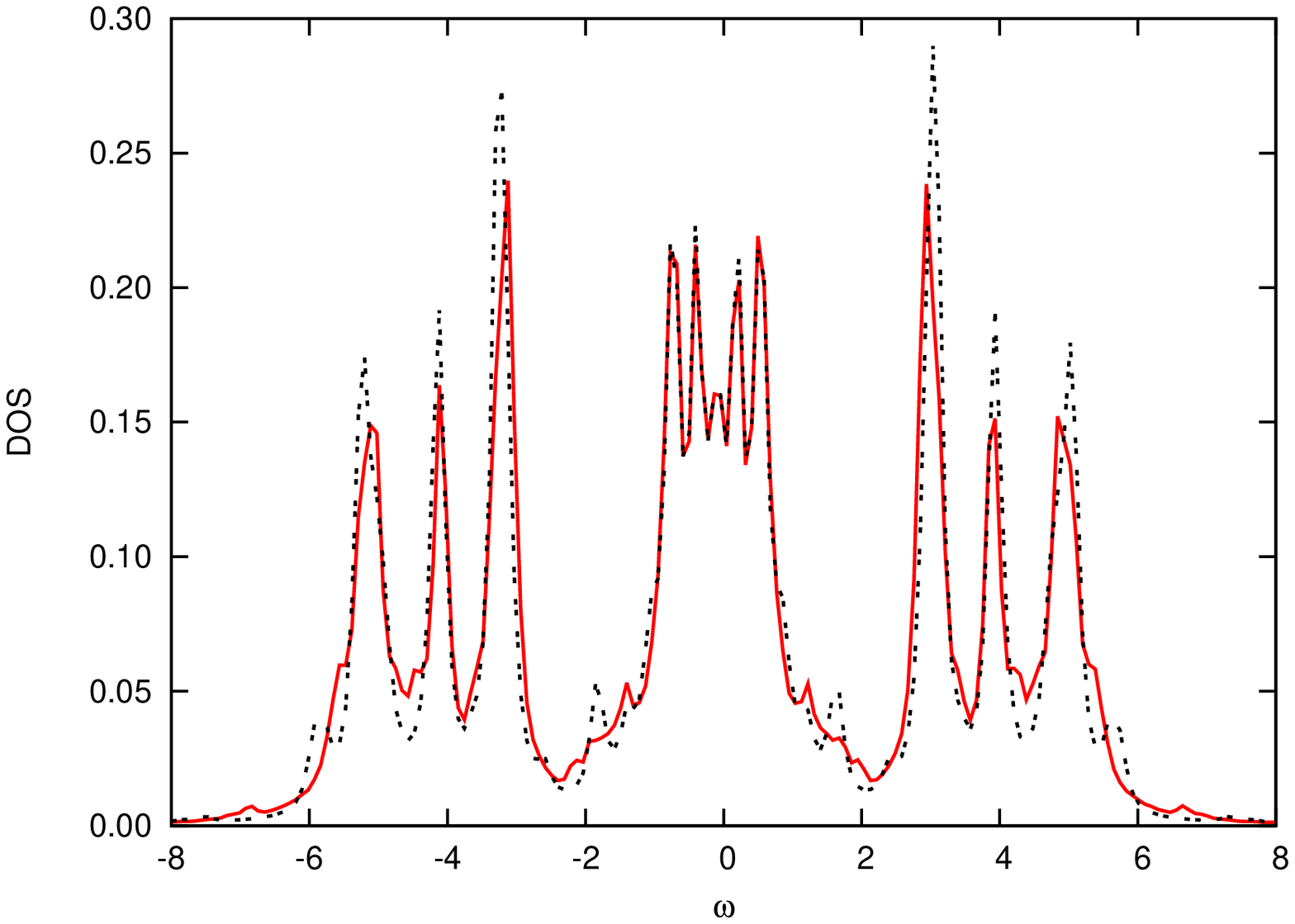} \\
 { (c)} \includegraphics[angle=270,width=0.95\columnwidth]{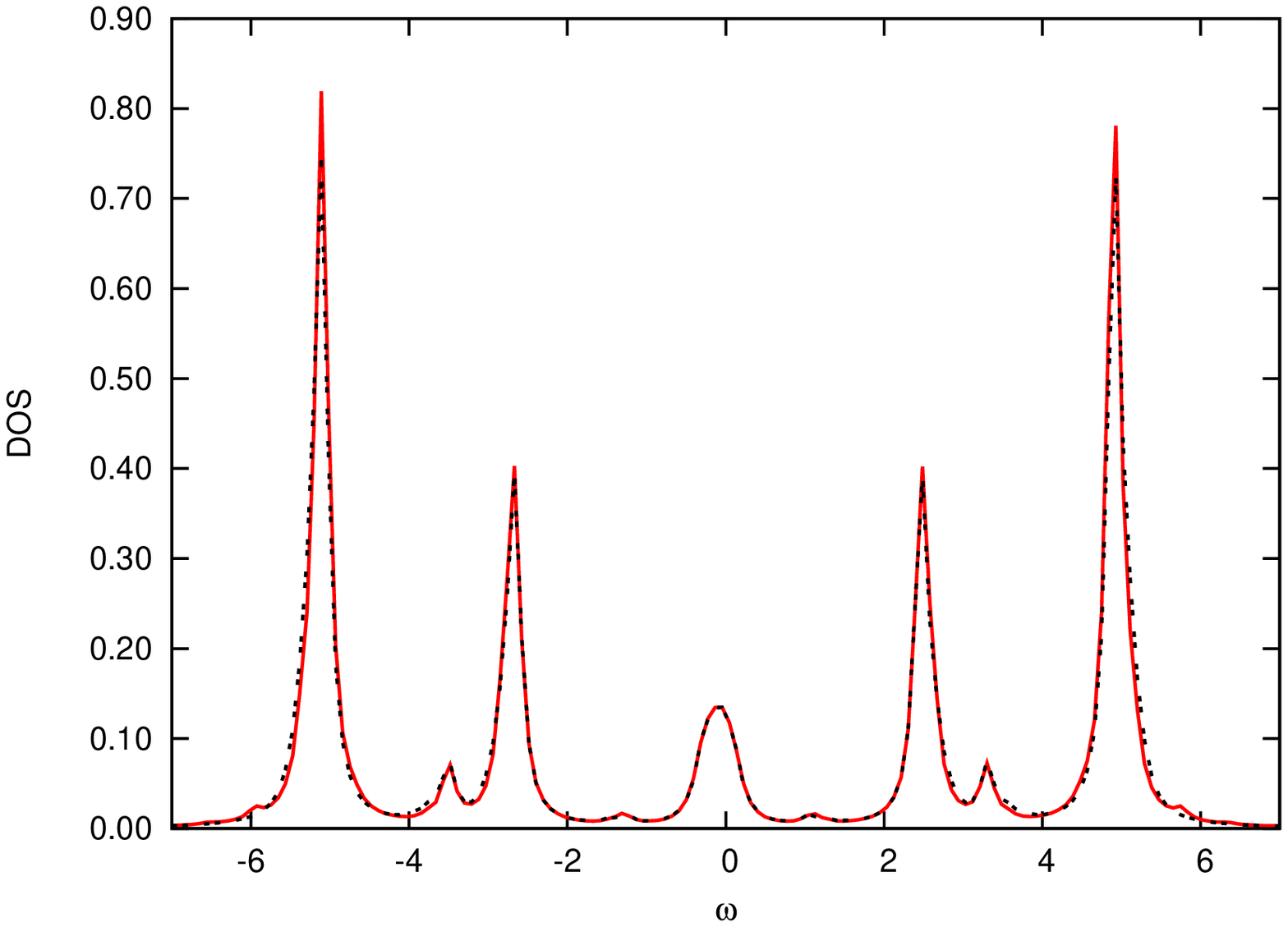} &
 { (d)} \includegraphics[angle=270,width=0.95\columnwidth]{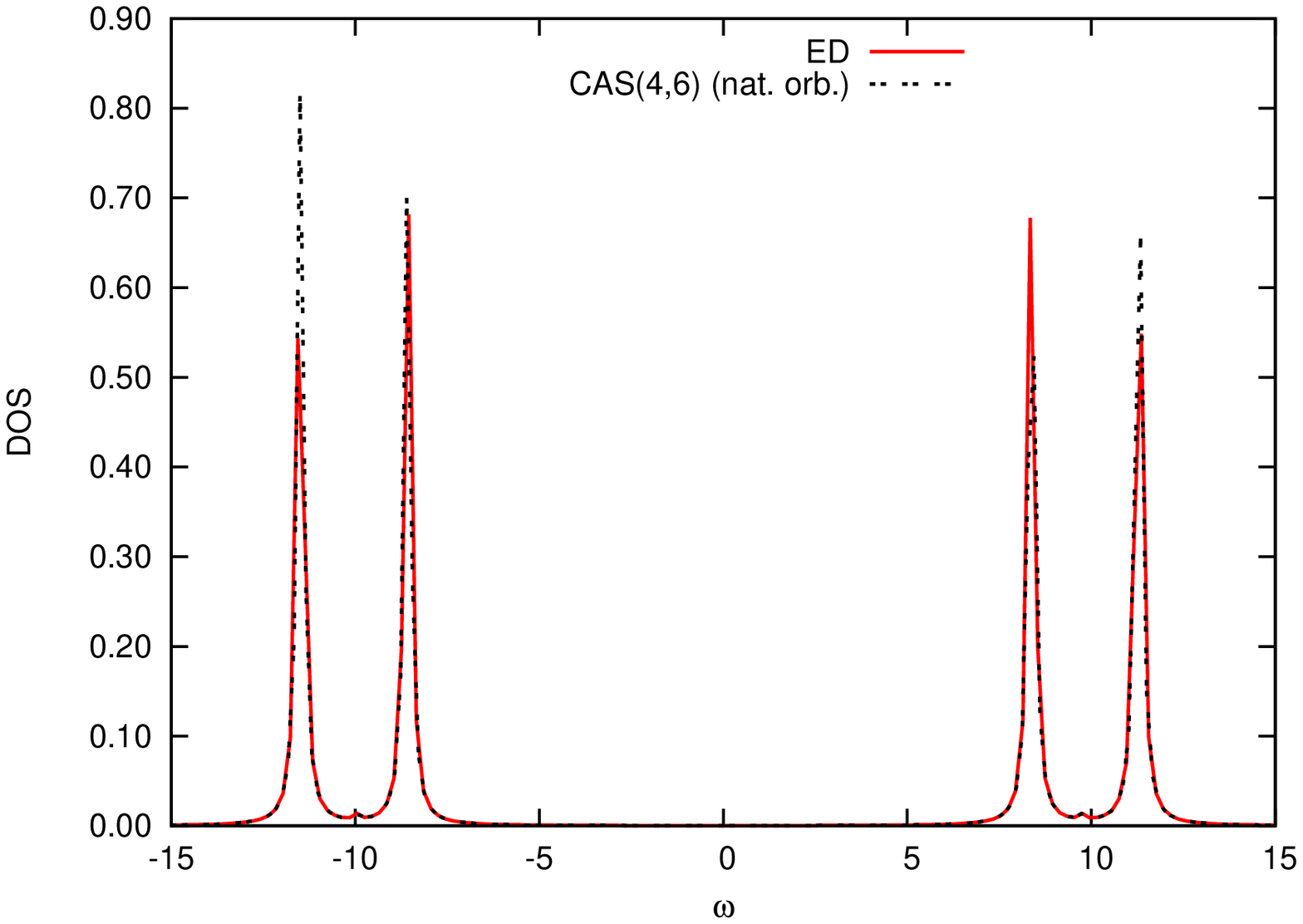} \\
\end{tabular}
\caption{Single site DMFT approximation to the 1D Hubbard model at half filling, using $11$ bath orbitals: 
Spectral function (DOS) $A(\omega)=-\frac{1}{\pi}\text{Im} G(\omega)$
Comparison between CAS(4,4) in the natural orbital basis (black dots) and ED (solid line).
(a) $U/t=4$, (b) $U/t=6$, (c) $U/t=8$, (d) $U/t=20$.}
\label{fig:natorb}
\end{center}
\end{figure*}

As discussed in section \ref{subsec:orbital}, the accuracy of the CI expansions
can be improved by working in the natural orbital basis. Examining
the ED calculations at half-filling
 we find that across the range of different $U/t$  only 
$4$ natural orbitals have occupancies appreciably
different from 0 and 2. Consequently, we choose these 4 active orbitals for an active space
calculation in the natural orbital basis.
 In Fig.~\ref{fig:natorb}, we show the spectral functions
at half-filling using the CAS(4,4) approximation,  in the natural orbital basis of the ED calculation. Note that the
 CAS(4,4) wavefunction involves \emph{ only 16 determinants} but the
spectral functions are still remarkably similar to the ED spectral functions. In fact, they are of similar quality to the CAS(2,2)CISD spectral functions (also in the natural orbital basis). This demonstrates the compactness of the natural orbital description.

\subsubsection{away from particle-hole symmetry}
\begin{figure*}[htb]
\begin{center}
\begin{tabular}{cc}
 { (a)} \includegraphics[angle=270,width=0.95\columnwidth]{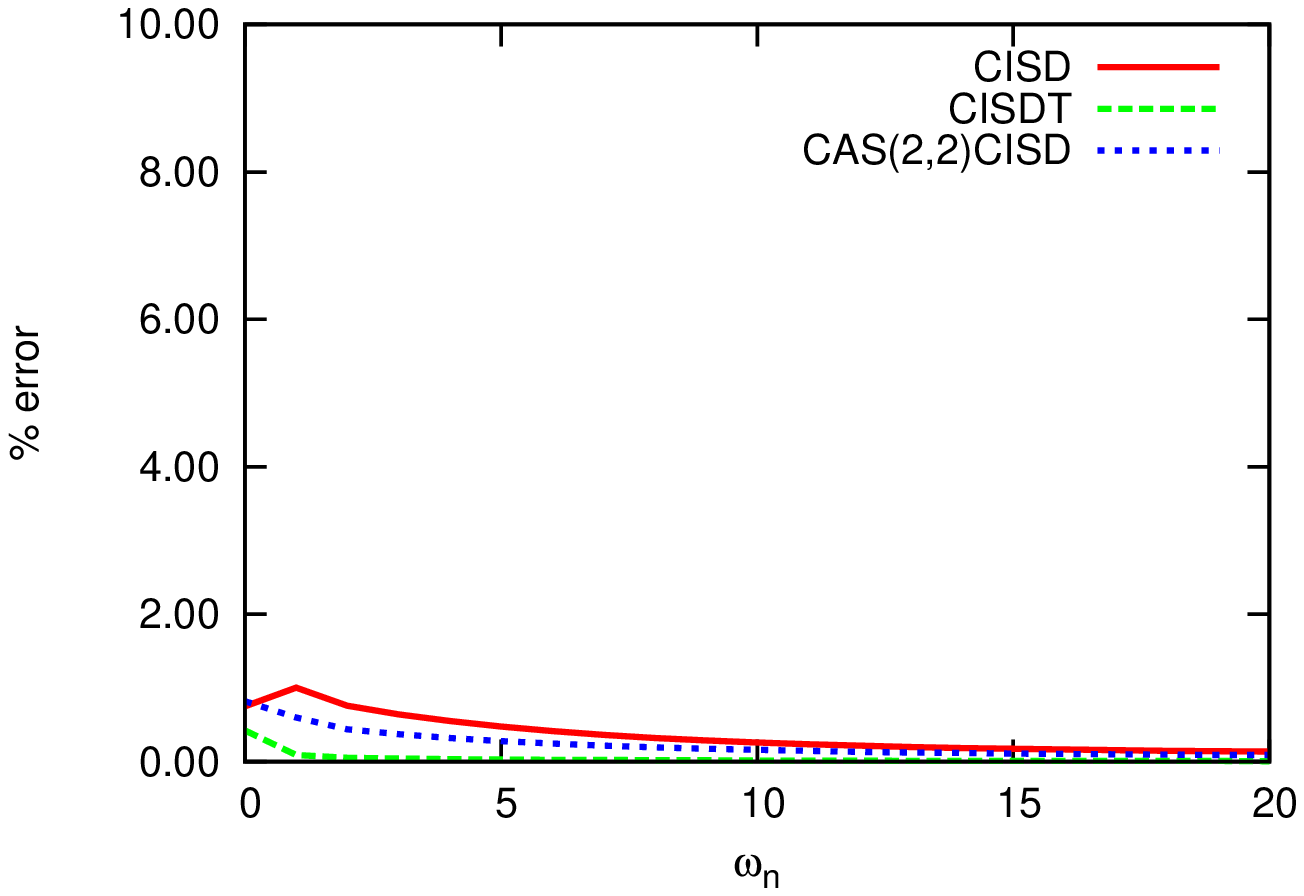} &
 { (b)} \includegraphics[angle=270,width=0.95\columnwidth]{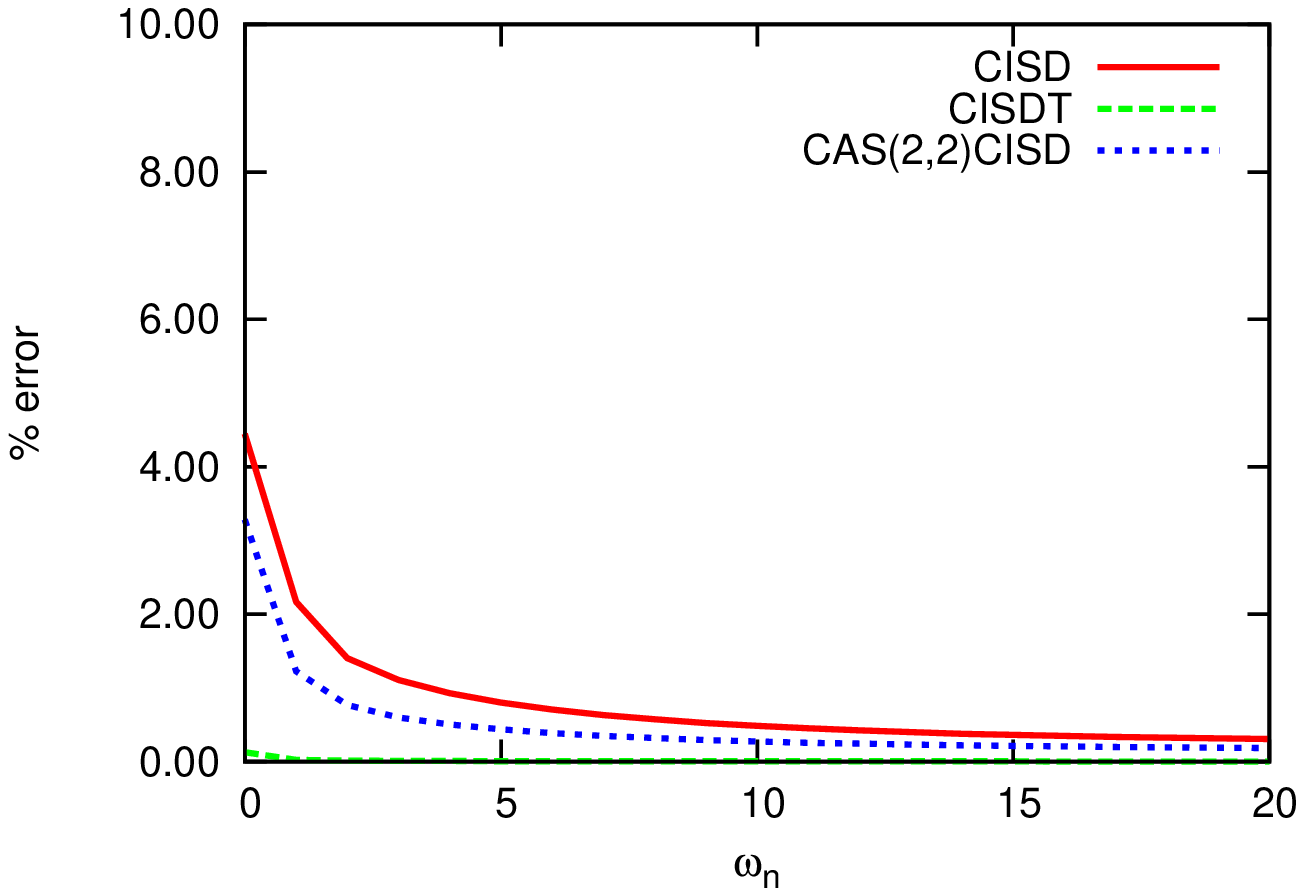} \\
 { (c)} \includegraphics[angle=270,width=0.95\columnwidth]{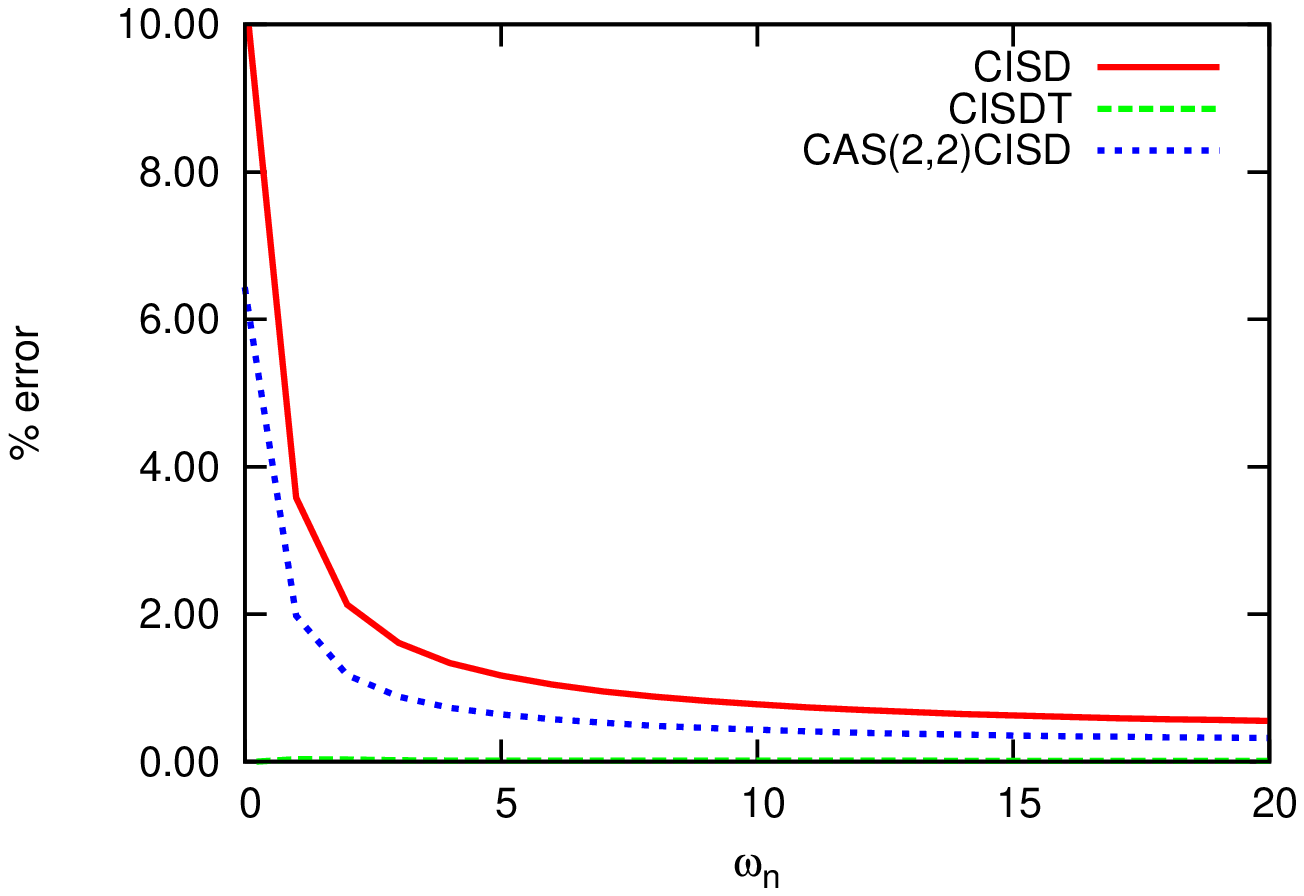} &
 { (d)} \includegraphics[angle=270,width=0.95\columnwidth]{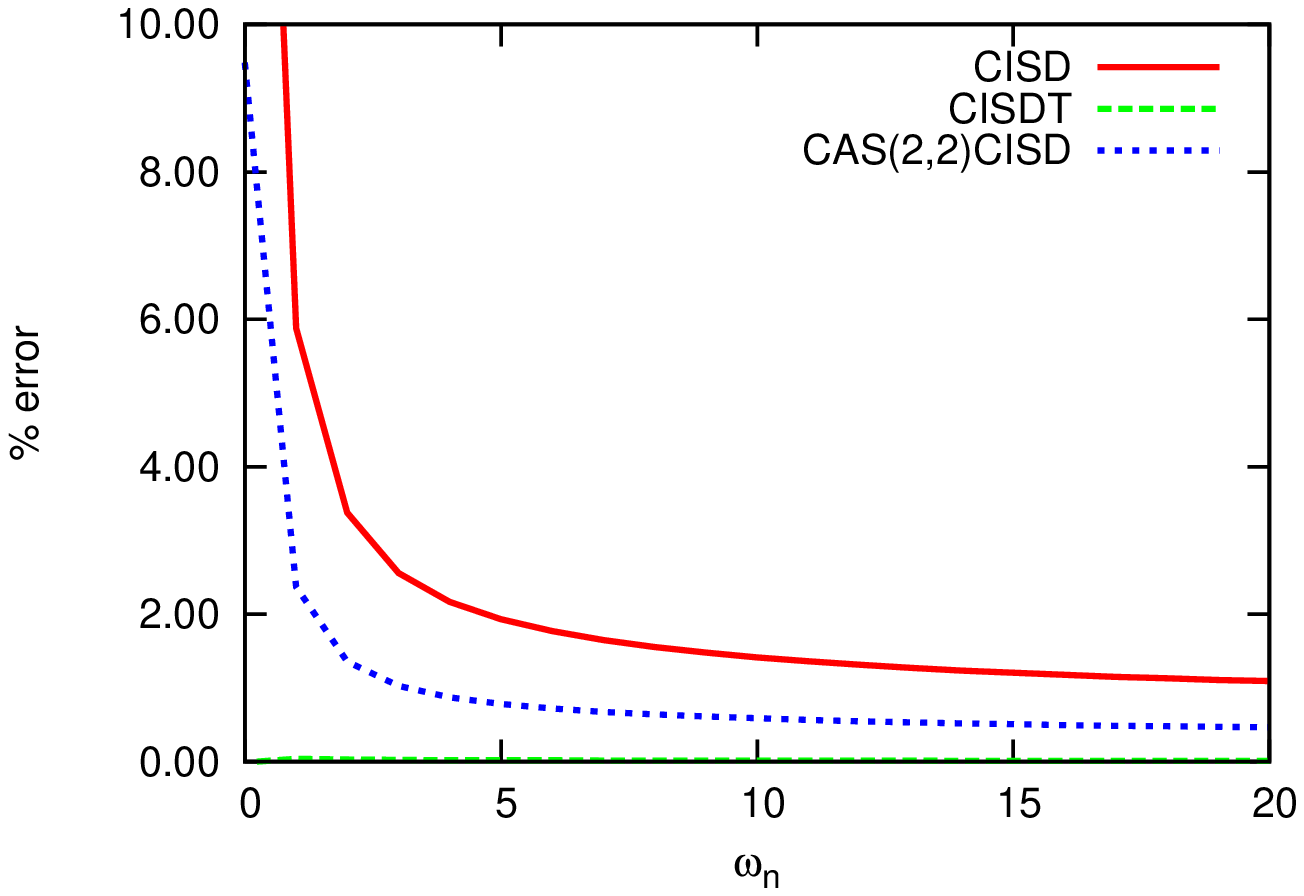} \\
\end{tabular}
\caption{Single site DMFT approximation to the 1D Hubbard model away from half filling, using $11$ bath orbitals:
Percent error in the imaginary part of the self-energy $\text{Im}\Sigma(i\omega_n)$ (relative to ED) for CISD (solid lines, red online), CISDT (dashed lines, green online), and CAS(2,2)CISD (dotted lines).
(a) $5\%$ doping, (b) $10\%$ doping, (c) $15\%$ doping, (d) $30\%$ doping. 
$\omega_n=(2n+1)\pi/\beta, \beta t=20$.
}
\label{fig:1D_away_hf}
\end{center}
\end{figure*}

We next consider the 1D Hubbard model away from half-filling. The corresponding imaginary parts of the self-energies,
for $U/t=6$ and dopings of $5\%-30\%$, are shown in Fig.~\ref{fig:1D_away_hf} for CISD, CISDT, and CAS(2,2)CISD. To better
illustrate the differences between the methods, here we plot the percentage error in the imaginary part of the self-energies, relative to ED. While
all the CI approximations yield qualitatively reasonable self-energies, we see that when we include triple excitations,
the errors become significantly less than $1\%$. 
This is consistent with our expectation that away from half filling, the wave function of this model becomes more single-determinantal and therefore it is more advantageous to base the description on a single reference determinant (in this case the HF determinant) than to include multiple determinant reference wave functions as in CAS(2,2)CISD.

\subsection{4-site cellular DMFT approximation to the 2D Hubbard model}

We now turn to cluster dynamical mean field theory, and in particular the 
2$\times$2 cellular DMFT approximation\cite{Kotliar01,Maier05} of the 2D Hubbard model. We begin with a 12 orbital quantum impurity model using
 8 bath orbitals, corresponding to 2 bath orbitals per impurity site, a model that
has been extensively studied in previous ED calculations.\cite{Civelli05,Kancharla08,Liebsch08,Liebsch09}
In common with these studies, we use the 4-fold
symmetry of the $2\times 2$ cluster and calculate the Green's function and self-energies
in the symmetry adapted basis of the cluster. In this basis, the Green's functions and self-energies become diagonal.
Labeling the sites of the 2$\times$2 cluster as $1\equiv(0,0)$, $2\equiv(1,0)$, $3\equiv(0,1)$, $4\equiv(1,1)$, 
the symmetry orbitals $\Gamma$, $M$, $X$ (doubly degenerate) are given by
\begin{align}
\psi_\Gamma &= \frac{1}{2}(\phi_{1}+\phi_{2}+\phi_{3}+\phi_{4})\\
\phi_M &=\frac{1}{2}(\phi_{1}-\phi_{2}-\phi_{3}+\phi_{4})\\
\phi_X &=\frac{1}{2}(\phi_{1}+\phi_{2}-\phi_{3}-\phi_{4})\label{Xself}\\
\phi_{X'} &=\frac{1}{2}(\phi_{1}-\phi_{2}+\phi_{3}-\phi_{4})
\end{align}
The symmetry orbitals $\phi_X$ and $\phi_{X'}$ form a degenerate pair. For a detailed description of this model see e.g.~Refs.~\onlinecite{Liebsch08,Liebsch09}.

We carried out CISD, CISDT, CAS(2,2)CISD, CAS(2,2)CISDT,  CAS(2,2)CISDTQ, and ED calculations of the spectral
functions and self-energies at half-filling. The CI calculations were carried out in the natural orbital basis of
a CAS(2,2)CISD calculation in the Hartree-Fock basis. 

\begin{figure*}[htb]
\begin{center}
\begin{tabular}{cc}
 { (a)} \includegraphics[angle=270,width=0.95\columnwidth]{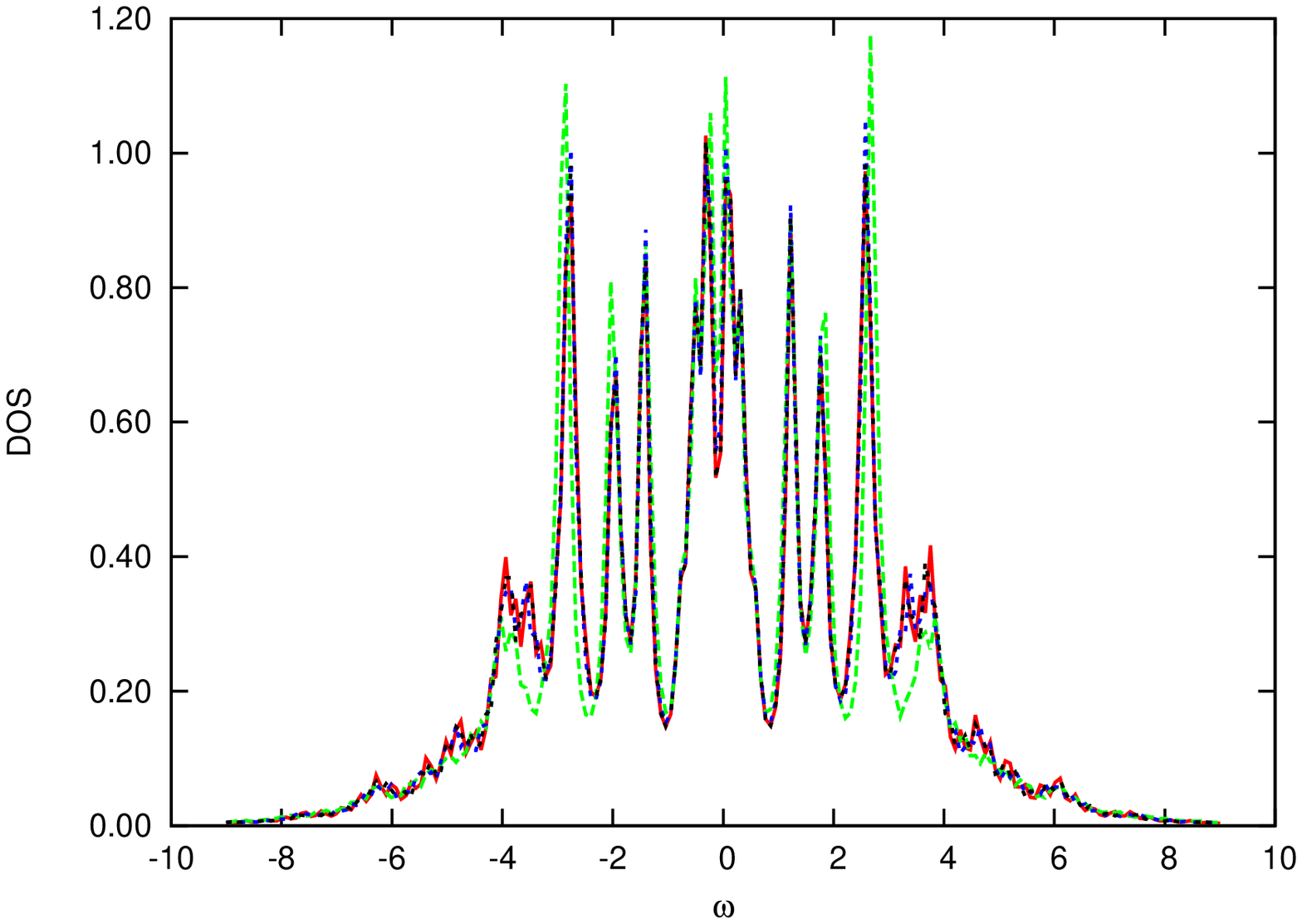} &
 { (b)} \includegraphics[angle=270,width=0.95\columnwidth]{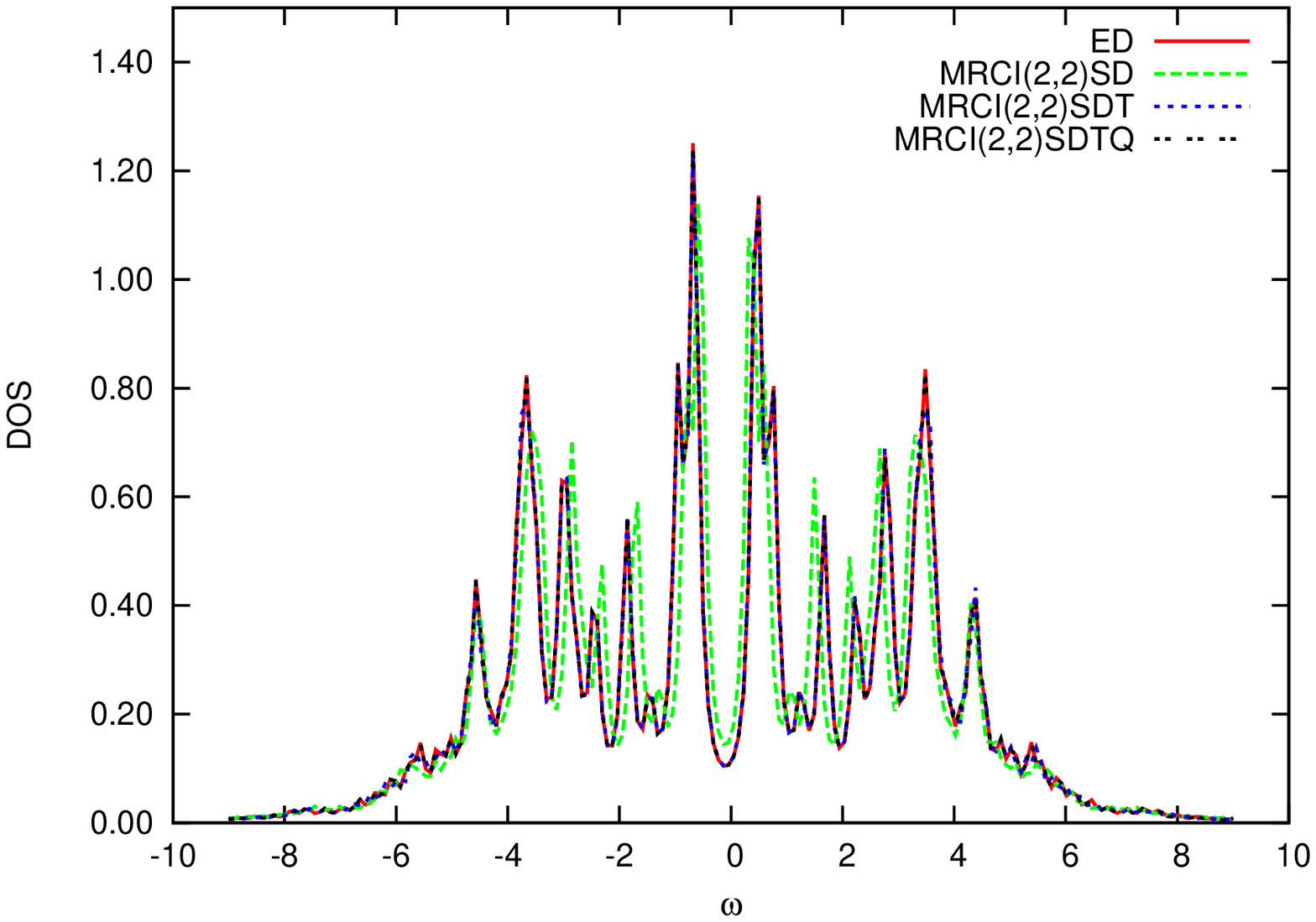} \\
 { (c)} \includegraphics[angle=270,width=0.95\columnwidth]{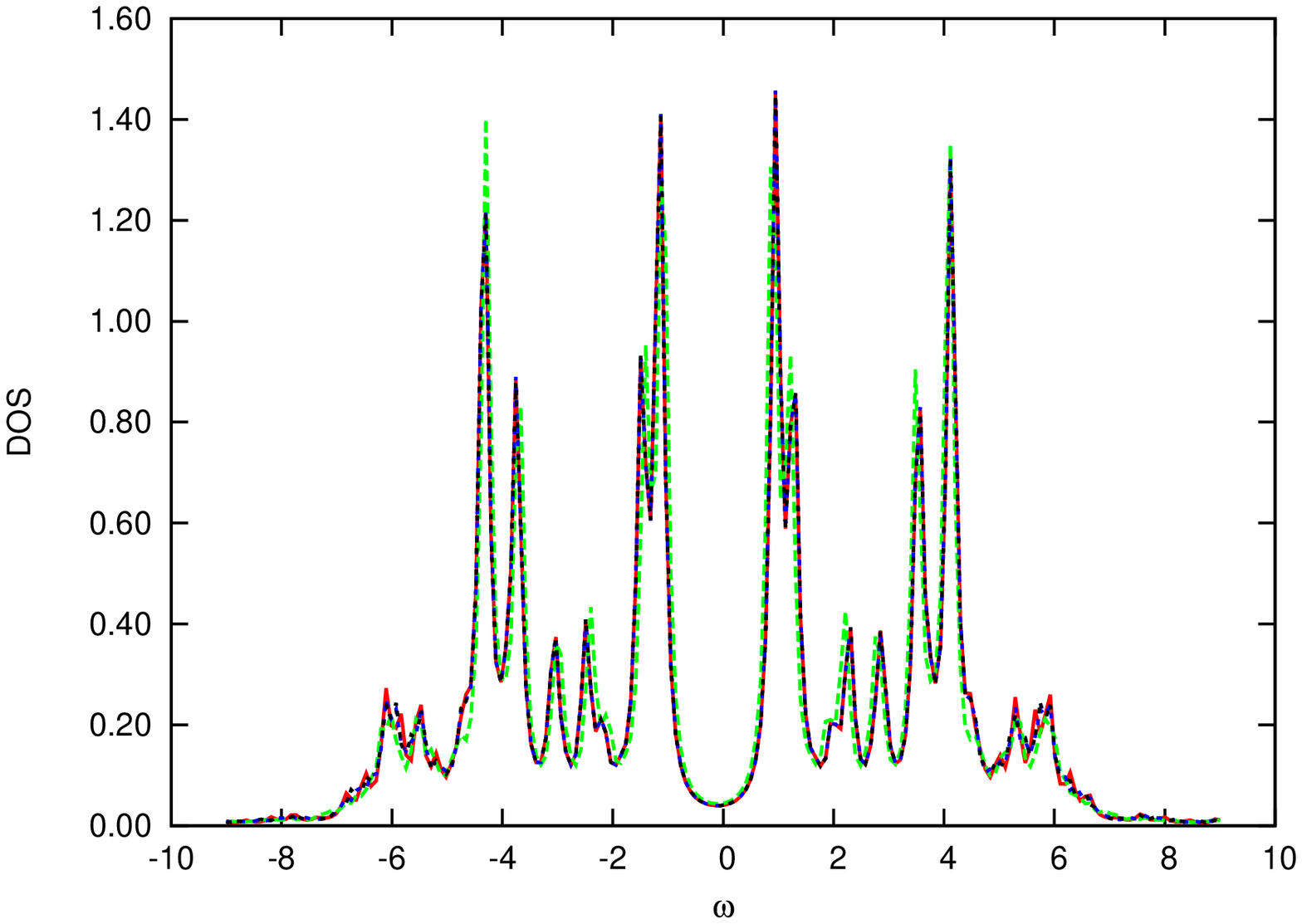} &
 { (d)} \includegraphics[angle=270,width=0.95\columnwidth]{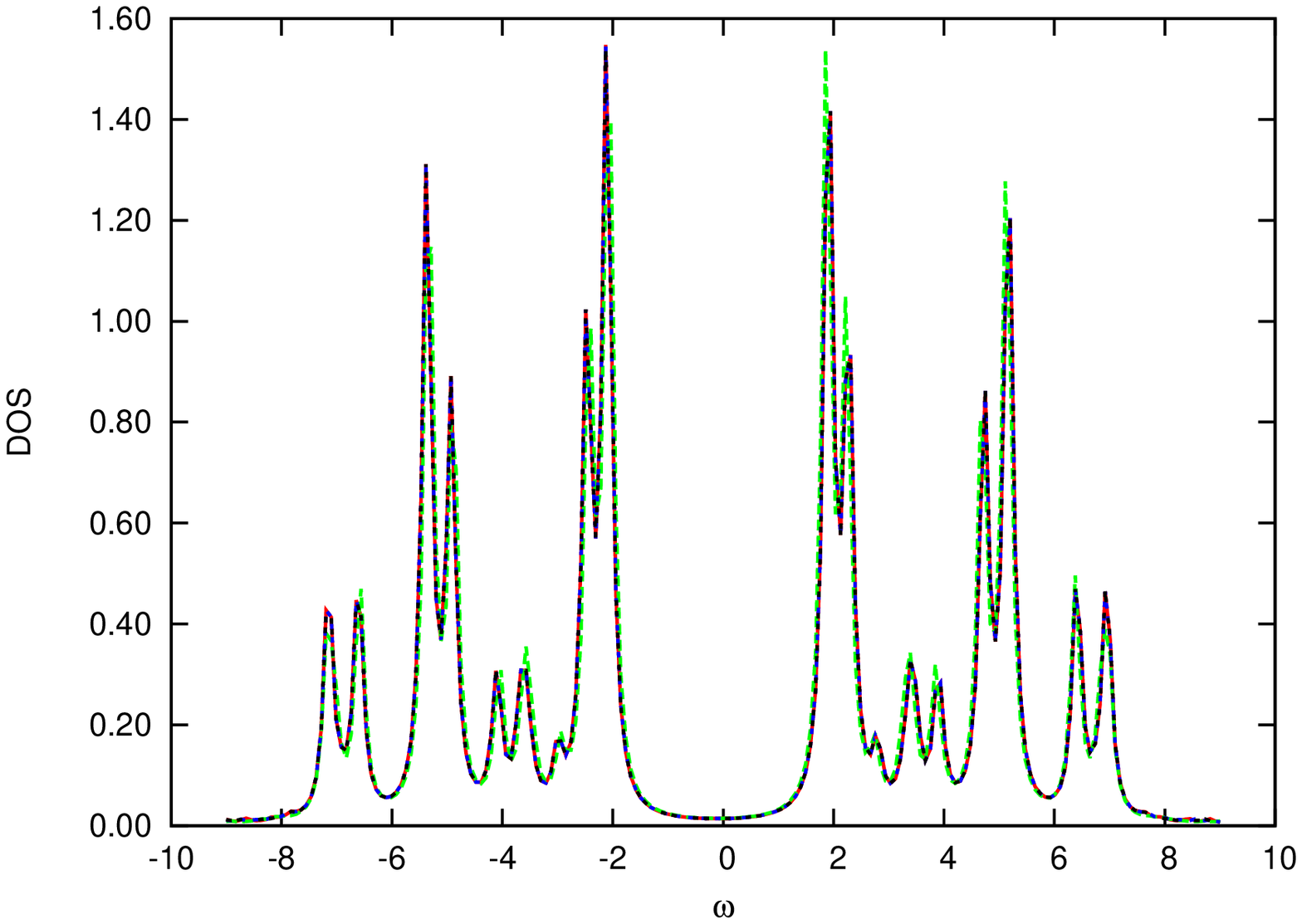} \\
\end{tabular}
\caption{Cellular dynamical mean field approximation to the $2D$ Hubbard model at half filling on a $2\times2$ cluster using $8$ bath orbitals:
Local spectral function (DOS) $A(\omega)=-\frac{1}{\pi}\text{Tr}\text{Im} G(\omega)$. For a description of the methods see text.
(a) $U/t=4$, (b) $U/t=5$, (c) $U/t=6$, (d) $U/t=8$.}
\label{fig:2D_dos}
\end{center}
\end{figure*}

\begin{figure*}[htb]
\begin{center}
\begin{tabular}{cc}
 { (a)} \includegraphics[angle=270,width=0.95\columnwidth]{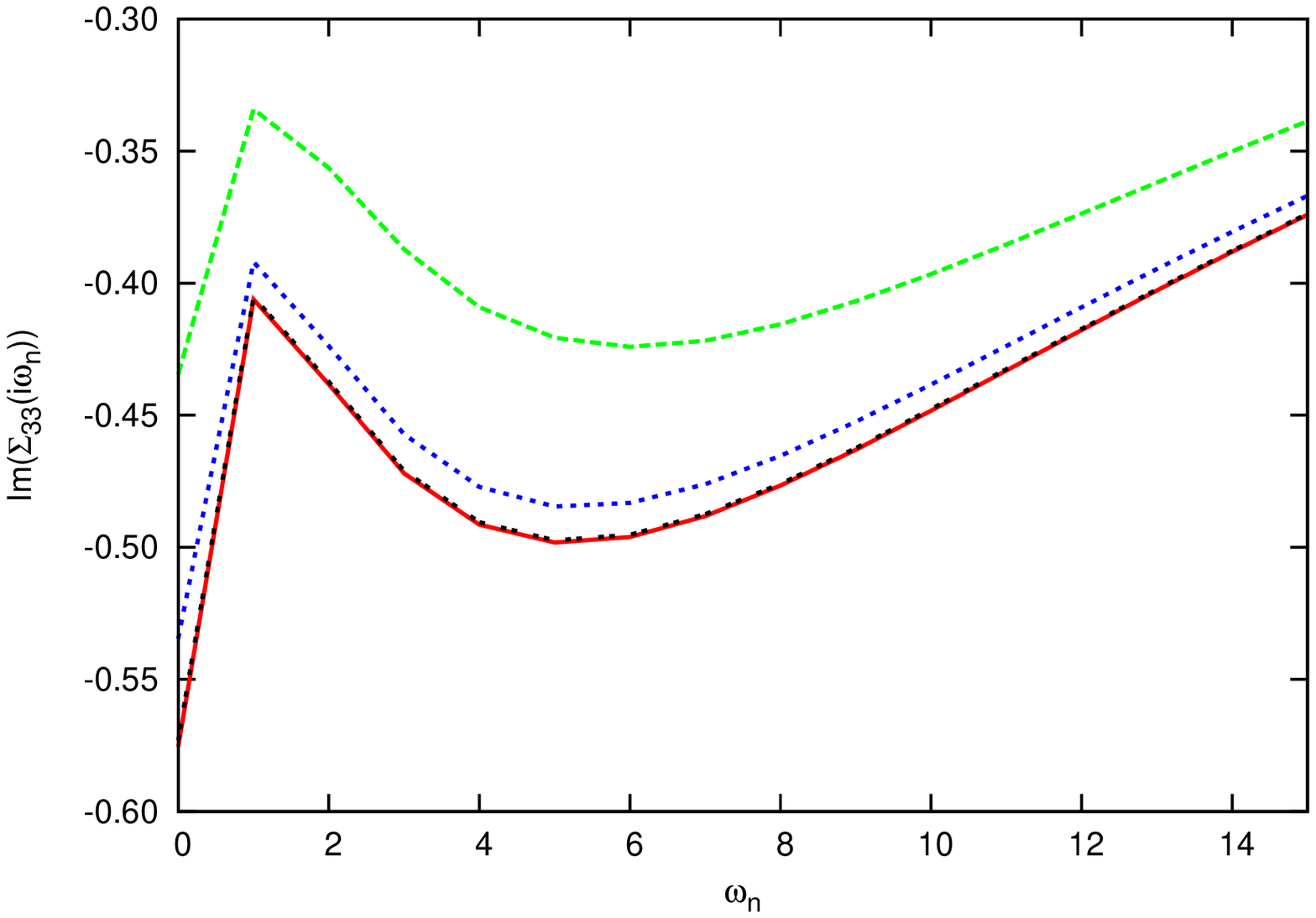} &
 { (b)} \includegraphics[angle=270,width=0.95\columnwidth]{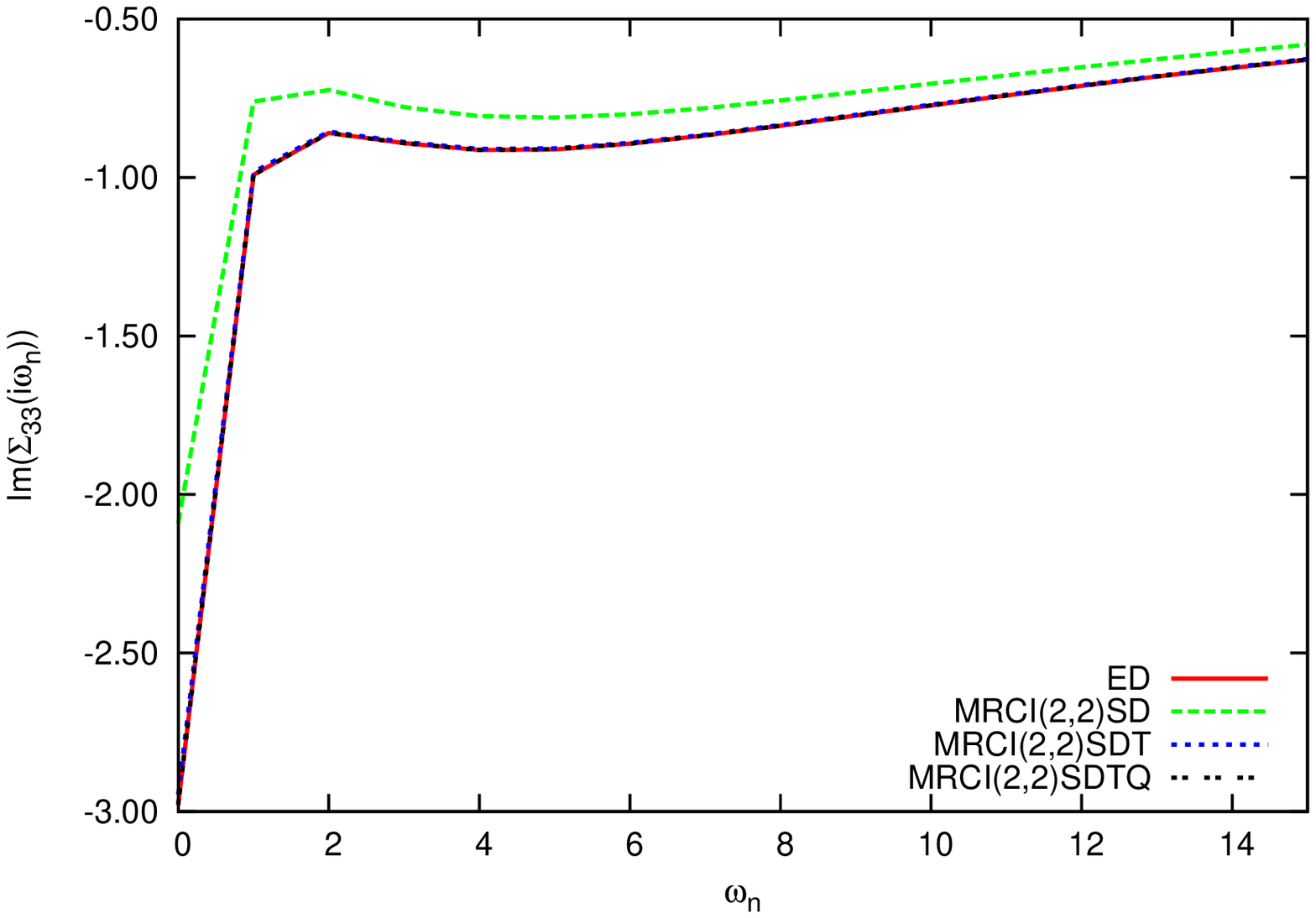} \\
 { (c)} \includegraphics[angle=270,width=0.95\columnwidth]{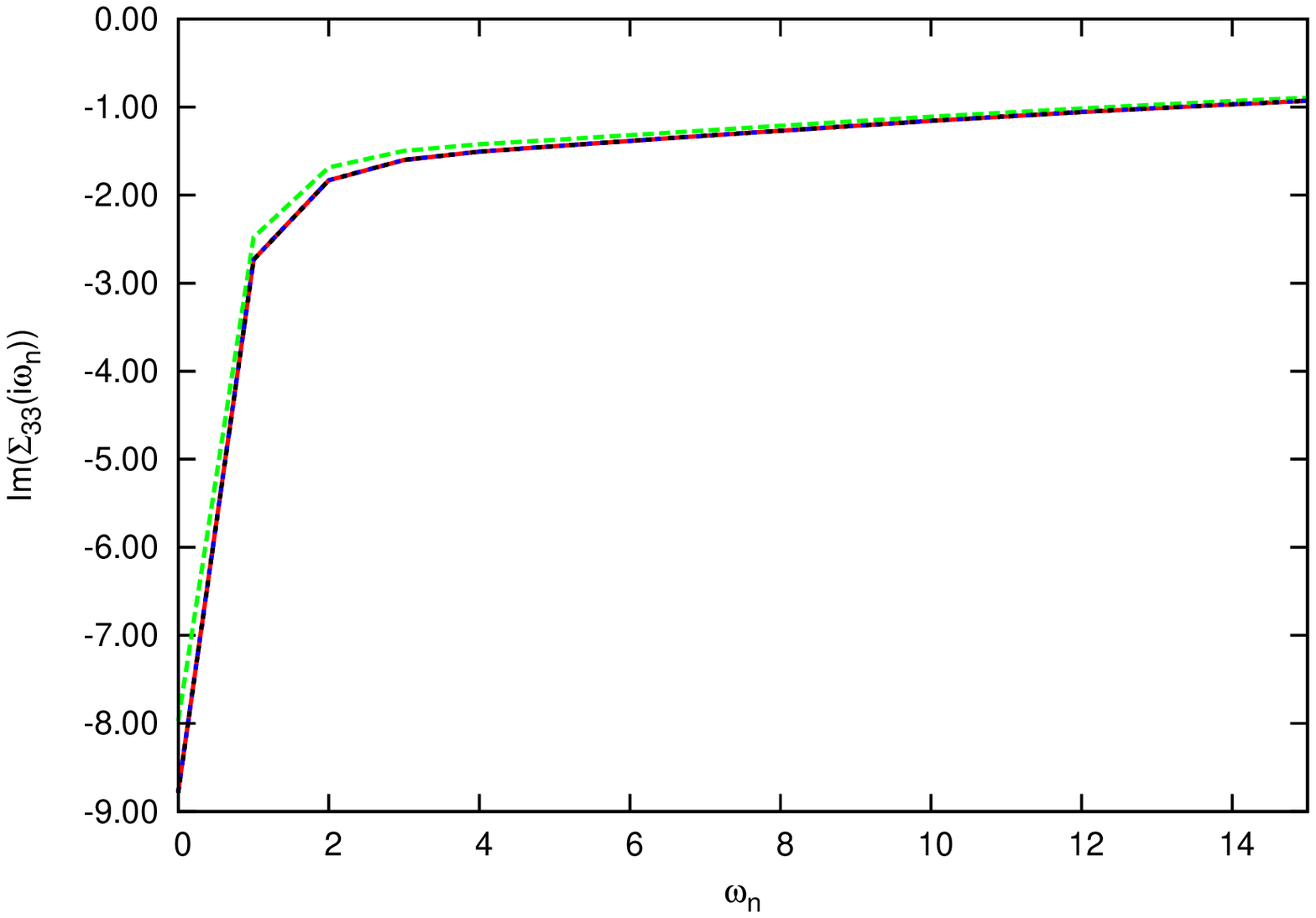} &
 { (d)} \includegraphics[angle=270,width=0.95\columnwidth]{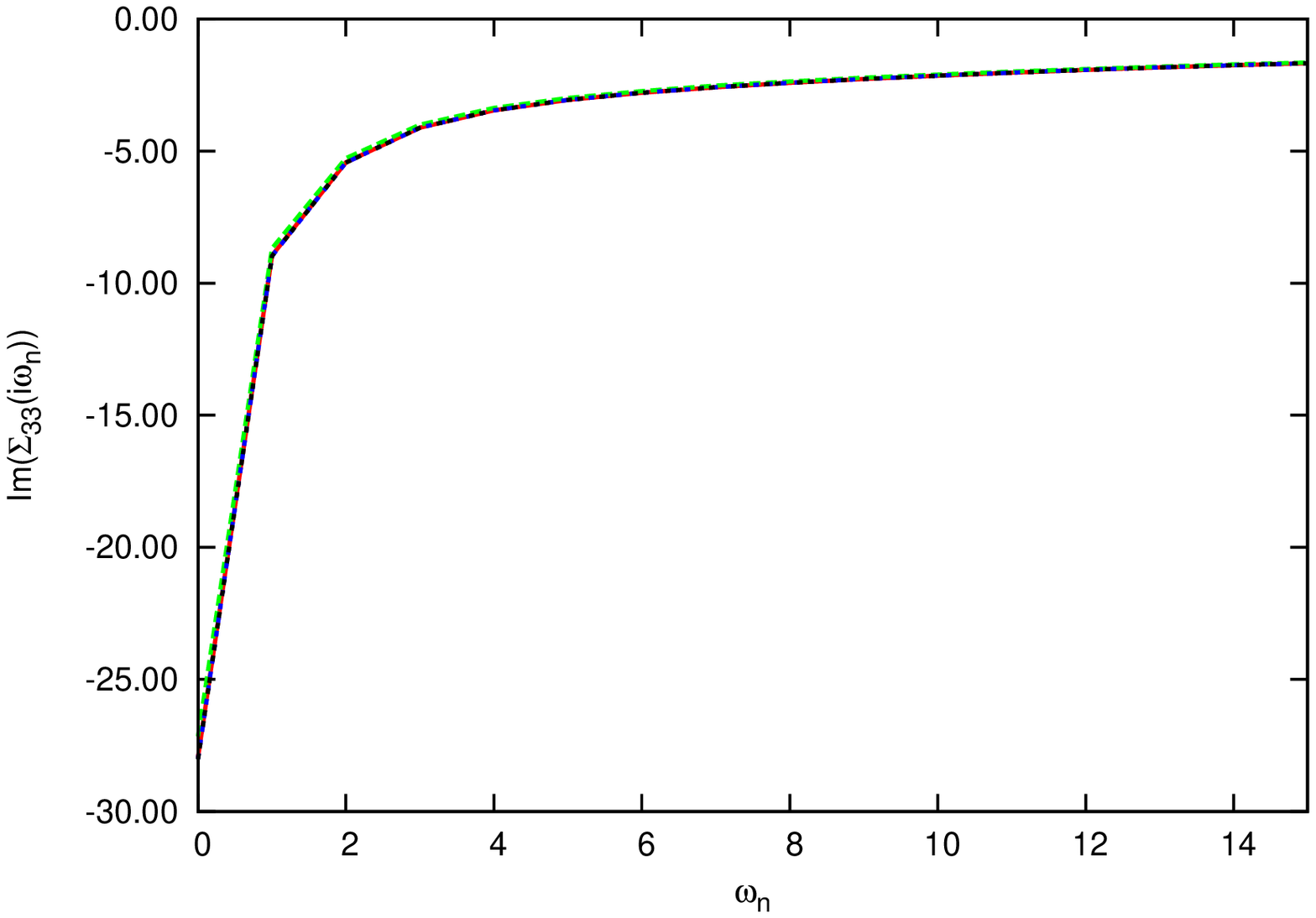} \\
\end{tabular}
\caption{Cellular dynamical mean field approximation to the $2D$ Hubbard model at half filling on a $2\times2$ cluster using $8$ bath orbitals:
Imaginary part of the self-energy $\text{Im} \Sigma_{33}(i\omega_n)$. For a description of the methods see text.
(a) $U/t=4$, (b) $U/t=5$, (c) $U/t=6$, (d) $U/t=8$.
$\omega_n=(2n+1)\pi/\beta, \beta t=12.5$.}
\label{fig:2D_se}
\end{center}
\end{figure*}

The local spectral functions $-\frac{1}{4\pi}\text{Tr} \text{Im} G(\omega)$ are shown in Fig.~\ref{fig:2D_dos}.
The imaginary parts of the $X$ self-energy, $\text{Im}\Sigma_X(\omega)$, corresponding to Eq.~\ref{Xself}, are shown in  Fig.~\ref{fig:2D_se}. Similar to the 1D case, we find good agreement
between all the CI methods and ED for all studied values of $U/t$, although there are some visible differences
between CAS(2,2)CISD and ED. Once triple and higher excitations are included, however, the spectral functions
become indistinguishable to the eye. The same conclusion can be drawn from analyzing the self-energies. 
While CAS(2,2)CISD is qualitatively similar to ED, the self-energy for $U/t=4$ is shifted from the ED self-energy,
with the errors largest at small frequencies. Once triples are included, the agreement becomes much better, and
with quadruples the self-energy is indistinguishable from that of ED. If we consider CAS(2,2)CISDT as yielding quantitative
agreement, then this is achieved using $49644$ determinants in the CI expansion, or only about $6\%$ of the ED determinantal space.

In this model, we find that the most difficult values of $U/t$ to achieve  agreement between the CI methods and ED
are for $U/t=4$ and $U/t=5$. Here, the form of the self-energy is that of a correlated Fermi liquid with a large effective mass. 
This behavior appears in the vicinity of the first-order cluster DMFT metal-insulator transition which, in CT-QMC simulations, is near $U/t=5.4$.\cite{Park08plaquette,Sordi10}  
(Note that in this pseudogap region, ED calculations can actually
converge to two different correlated metallic solutions, depending
on the initial guess for the bath parametrization, a
feature which is repeated in the CI calculations. We have chosen to present the more insulating solution in  Fig.~\ref{fig:2D_se}). 

\begin{figure}[htb]
\begin{center}
\includegraphics[angle=270,width=0.95\columnwidth]{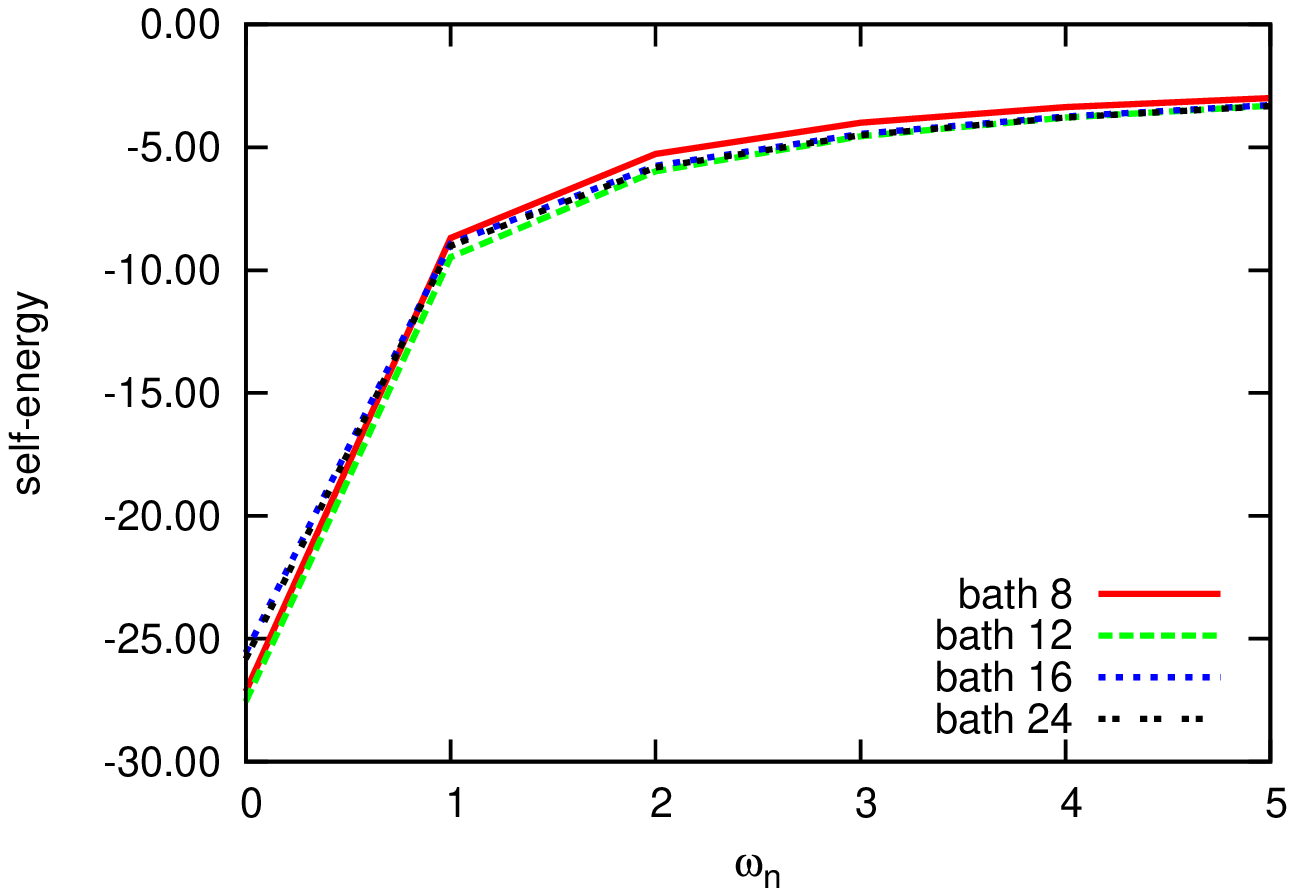}
\caption{Cellular dynamical mean field approximation to the $2D$ Hubbard model at half filling on a $2\times2$ cluster for a range of bath sites:
Imaginary part of the self-energy $\text{Im} \Sigma_{33}(i\omega_n)$ 
The CI method used was CAS(2,2)CISD in the natural orbital basis of CAS(2,2)CISD.}
\label{fig:convergence}
\end{center}
\end{figure}

We now briefly turn to some calculations on this model which cannot be performed using ED. An essential
weakness of ED (and CI) solvers in the DMFT context is the need to parametrize the bath
using a finite number of bath orbitals. If the number of bath orbitals is too small, the
resolution of the spectral function and other quantities is very low, and furthermore, artifacts
can appear in the ED calculations due to a large fitting error at low frequencies.\cite{Koch08,Liebsch09,Senechal10} CI approximations, however, allow us to treat larger numbers
of orbitals, and thus potentially alleviate the bath parametrization problem by allowing us to
use a sufficient number of bath orbitals. We now demonstrate this for the $2\times 2$ cluster. In
Fig. \ref{fig:convergence} we plot the self-energies for $U/t=8$ from CAS(2,2)CISD calculations (using  the CAS(2,2)CISD natural orbital basis) 
and for $8, 16,$ and $24$ bath orbitals. For $12$ bath orbitals we used CAS(4,4)CISD rather than CAS(2,2)CISD for technical reasons due to the degeneracy of the reference wave function. The largest calculation with 24 bath orbitals (or a total of 28 orbitals in
the impurity model) is roughly twice the size of what can be treated with ED. Our studies confirm that  convergence in this model
is achieved fairly rapidly, but that there are nonetheless quantitative differences
between the standard 8 bath orbital parametrization and larger bath representations, particularly
for small frequencies. The $16$ bath orbital and $24$ bath orbital parametrizations are indistinguishable, indicating that full convergence has been reached.
We have also carried out calculations for other values of $U/t$, where we observe similar convergence behavior.
The convergence of the bath parametrization appears slower for $U/t=5$ and $U/t=6$, which may once again be related to the proximity to a metal-insulator transition.

\section{Conclusions}\label{sec:conclusions}

In the current work we have described how configuration interaction (CI) approximations
to exact diagonalization (ED) can be used as solvers for quantum impurity models,
such as those encountered in dynamical mean-field theory (DMFT). CI solvers
form a controlled hierarchy of polynomial cost approximations that
retain the main advantages of ED, such as the ability to treat
general interactions and obtain real spectral information. As we have demonstrated in this work,
 the convergence of the CI hierarchy is sufficiently rapid that in many cases, they almost
exactly approximate the ED results, at a small fraction of the cost. This is true even in ``difficult'', ``strongly correlated'' regimes,
such as the pseudogap regime of the $2\times 2$ cluster DMFT of the Hubbard model. In addition, this great increase
in computational efficiency potentially allows us to treat considerably larger quantum impurity
models than have been considered in ED. In this work we used this ability to demonstrate
bath convergence for the $2\times 2$ cluster DMFT of the Hubbard model in a calculation with $28$
orbitals, for a case where previously only $12$ orbitals (Lanczos) were accessible. 

Here we have focused on well studied DMFT problems in order to benchmark the CI approximations.
In future work,  we plan to apply these CI approximations to study problems where existing solvers
have difficulties. Some of these include impurity models with a large number of orbitals and with
general interactions and off-diagonal hybridizations, for which CT-QMC methods encounter a severe sign problem. Another interesting direction to explore will be
to examine more sophisticated quantum chemistry approximations to ED. For example, for weak interactions,
coupled cluster approximations are known to be far superior to configuration interaction approximations
for a given computational cost. These and other directions are currently being pursued.

\acknowledgments{Dominika Zgid acknowledges helpful discussions with A.J.~Millis, D.R.~Reichman, A.I.~Lichtenstein, L.~de~Medici, and A.~Liebsch. Dominika Zgid and Garnet Kin-Lic Chan acknowledge support from the Department of Energy, Office of Science,
through Award DE-FG02-07ER46432. Emanuel Gull was partially supported by NSF-DMR-1006282.}

\appendix
\section{Three-orbital model with rotationally invariant interactions}\label{sec:threeorb}
As a further application we present here results for a three-orbital model with general, rotationally invariant interactions.
Problems of this type have been notoriously difficult to solve, as quantum Monte Carlo impurity solvers for multi-orbital models are 
either limited to density-density interactions\cite{Hirsch86,Gull08_ctaux}
or suffer from a severe fermionic sign problem, even at half-filling.\cite{Sakai04,Rubtsov05,Sakai06}
So far, only the continuous-time hybridization expansion\cite{Werner06,Werner06Kondo} and exact diagonalization methods\cite{Liebsch12} 
have been able to access this regime, but the exponential scaling
of the local impurity Hilbert space size makes five- and seven-orbital systems inaccessible without severe truncations or fitting errors.\cite{GullRMP}

\begin{figure}[htb]
\begin{center}
\includegraphics[angle=270,width=0.95\columnwidth]{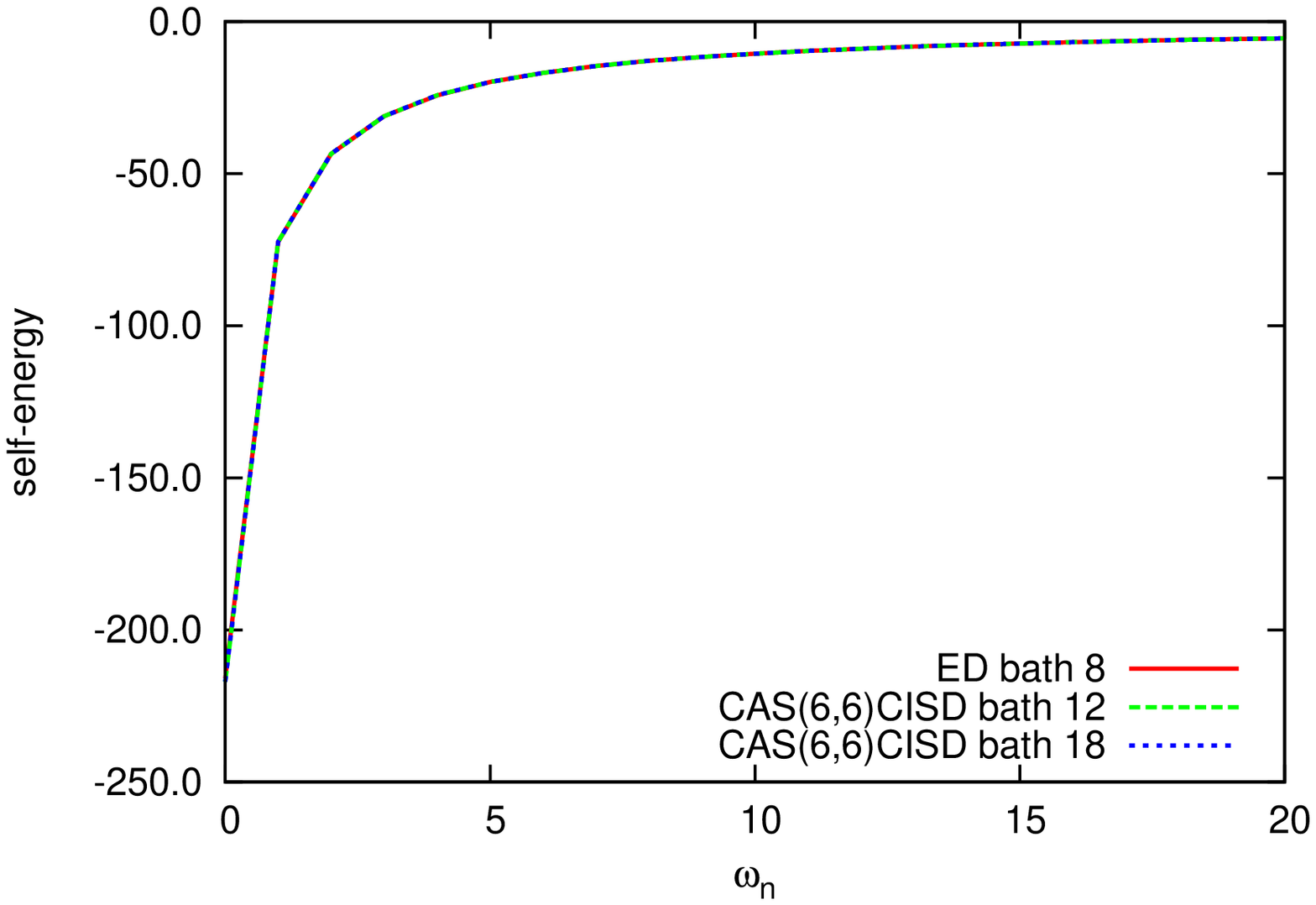}
\caption{Dynamical mean field approximation to the 3-orbital Hubbard model at half filling: Imaginary part of the self-energy $\text{Im}\Sigma(i\omega_n)$ for ED (solid lines, red online) with $12$ bath sites, CAS(6,6)CISD with $15$ bath sites, and CAS(6,6)CISD with $21$ bath sites. Differences are on the order of $0.5\%$. }
\label{fig:threeorb}
\end{center}
\end{figure}

The three orbital model with the Slater-Kanamori\cite{Mizokawa95,Imada98} form of the Hamiltonian,
\begin{align}
H_\text{loc}&=H_\text{SK}\equiv U\sum_an_{a\uparrow}n_{a\downarrow}+(U-2J)\sum_{a\neq b}n_{a\uparrow}n_{b\downarrow} 
\nonumber \\
&+(U-3J)\sum_{a> b,\sigma}n_{a\sigma}n_{b{\sigma}} \nonumber \\ 
&-J\sum_{a\neq b}\left(d^\dagger_{a\uparrow}d^\dagger_{a\downarrow}d_{b\uparrow}d_{b\downarrow} +d^\dagger_{a\uparrow}d^\dagger_{b\downarrow}d_{b\uparrow}d_{a\downarrow}\right),
\label{HSK}
\end{align}
on a Bethe lattice is a toy model that 
has been well studied with these methods\cite{Werner08nfl,Werner09OSMT,deMedici11,Liebsch12} and shows interesting spin-freezing behavior as a function of the Hund's coupling $J$.
We show in Fig.~\ref{fig:threeorb} the imaginary part of the self-energy at half filling (in the Mott insulating phase), for $U/t=12$ and $J/t=1$.
As in the case of the single- and four-orbital models, convergence to the ED solution for a fixed number of bath sites and convergence as a function of the number bath sites is observed, and we find no additional complications caused by the more general interaction structure.

\bibliographystyle{apsrev4-1}
\bibliography{refs}
\end{document}